\documentclass[aps,pra,reprint,groupedaddress]{revtex4-2}

\usepackage{graphicx}
\usepackage{amssymb,amsmath}
\usepackage{bm}
\usepackage{dcolumn}
\usepackage[pdftex,breaklinks,colorlinks,citecolor=blue,linkcolor=blue,urlcolor=blue]{hyperref}
\usepackage{xcolor}

\usepackage{mathrsfs}
\usepackage{enumerate}

\begin{document}

\title{Rigidity, separability, and cusp conditions of a wave function}

\author{Bo Gao}
\email[]{bo.gao@utoledo.edu}
\homepage[]{http://bgaowww.physics.utoledo.edu}
\affiliation{Department of Physics and Astronomy,
	Mailstop 111,
	University of Toledo,
	Toledo, Ohio 43606,
	USA}

\date{August 10, 2022}

\begin{abstract}

We introduce in quantum mechanics a concept of \textit{rigidity} and a concept of a \textit{pinned point} of a wave function. The concept of a pinned point is a generalization of a familiar concept in the description of a vibrating string, while the concept of rigidity is introduced to describe the sensitivity of a wave function to changes in energy, potential, and/or external perturbation. Through these concepts and their mathematical implications, we introduce and formulate cusp conditions and cusp functions as fundamental properties of an arbitrary $N$-body quantum system with $N\ge 2$, greatly expanding their relevance beyond the Coulombic systems. The theory provides rigorous constraints on an arbitrary $N$-body quantum system, specifically on its short-range pair correlation that is essential to a better understanding of strongly correlated systems. 
More broadly, the theory and the derivations presented here are part of a reconstruction of the mathematical and conceptual foundation of an $N$-body quantum theory, incorporating previously hidden properties and insights revealed through the concepts of rigidity and pinned points. It includes general analytic properties of a 2-body wave function versus energy, and their relations to cusp conditions and cusp functions. It includes a rigorous derivation and an understanding, in terms of an emergent length scale, of the 2-particle separability in an $(N>2)$-body quantum system and its relations to cusp conditions. It also includes a classification of quantum systems, both 2-body and $N$-body, based on the universal behaviors in their short-range correlation.
For systems of Coulombic interactions, the theory extends their cusp functions to infinite range and puts their understanding within a generalized Coulomb class covering all potentials with a behavior of $v(r)\overset{r\to 0}{\sim} \pm D_{\alpha}/r^{\alpha}$ with $\alpha<2$, including the cases of negative $\alpha$. 

\end{abstract}


\maketitle

\section{Introduction}

In the effort to understand the world around us as a quantum system of atoms and molecules, such as building a quantum theory for liquids and for interactions and reactions therein, we encounter two broad classes of difficulties. One has to do with the interaction, and the other has to do with the quantum treatment of strongly interacting particles, especially particles that can attract and bind. As put succinctly by Evans \textit{et al.}, ``\textit{most molecules are not like argon: The Lennard-Jones model has its limitations}'' \cite{Evans19}. Atomic and molecular interactions, from potentials \cite{Derevianko99,Samuelis00,Stone13} to scattering and reactions \cite{Bala16}, can already be complex. If they are so complex to be systemic-specific and to defy any universal behavior, there would be no meaningful $N$-atom or $N$-molecule theory beyond system-specific computation. Universal behavior in 2-body interaction is a prerequisite for a meaningful $N$-body theory. 

A recent development on interaction is that of a multichannel quantum-defect theory for anisotropic potentials (MQDTA) \cite{Gao20b}. With this progress, we can finally assert that atomic and molecular interactions in the intermediate and long range do in general fall into universal classes and can be treated systematically in a generalized quantum-defect theoretical (QDT) framework \cite{Gao20b} that builds upon the more traditional frameworks \cite{Seaton83,Greene79,Greene82,Greene85,Gao08a}. Thus the universal behaviors that have been found and verified in atomic and molecular interactions and reactions (see, e.g., Refs.~\cite{Gao98b,Flambaum99,Gao04b,Gao04c,Gao10a,Ospelkaus10,Gao10b,Gao11a}) are not accidental, nor fundamentally limited in scope. They are initial examples that can be extended to more atomic and molecular classes including those with complex anisotropic interactions, and be extended to cover greater ranges of energies through multiscale generalizations \cite{Gao16a,Hood20a}. Much further developments along those lines await, especially in the context of an $N$-body theory \cite{Gao03,Gao04a,Gao05b,KG06}, but it is clear conceptually \cite{Gao03,Gao04a} that universal behaviors and systematic theories should exist in $N$-atom or $N$-molecule systems at least for the intermediate and the long range correlations.

The other fundamental difficulty is the generic $N$-body difficulty for strongly interacting particles, especially particles that can attract and bind, which is true for basically all atoms and molecules in their ground electronic state. Compared to the substantial success of treating many electrons (see, e.g., Refs.~\cite{Burke12,Gruneis17}), which repel each other \cite{Gao17a} via Coulomb repulsion, our understanding of $N$-body quantum systems made of particles that can attract and bind has been limited to a few special cases such as He (see, e.g., \cite{Ceperley95,Leggett06}), with its unusually weak interaction, or systems of ultra-low densities at ultracold temperatures (see, e.g., Refs.~\cite{GPS08,BDZ08,Bohn17}). The lack of fundamental progress in this area over many decades is an indication that either something fundamental is missing, or there will be no true theory for such systems.

This work aims to show that a key missing component is at the fundamental level of quantum mechanics, in concepts that we will introduce as the concept of \textit{rigidity} and the concept of a \textit{pinned point}. These concepts and their mathematical implications lead to a reconstruction of the mathematical and conceptual foundation of an $N$-body theory that includes a general formulation of cusp conditions and their universal behaviors, far beyond that for the Coulomb potential \cite{Kato57,Pack66,Fournais05,Hattig12,Gruneis17}. It also includes a much more complete understanding of the analytic property of a 2-body wave function versus energy, a rigorous derivation and understanding of 2-particle separability in an $(N>2)$-body quantum system, and a classification of quantum systems, both 2-body and $N$-body, based on the universal behaviors in their short-range pair correlation. This reconstruction is a key step towards a quantum $N$-body theory of strongly interacting particles, especially particles that can attract and bind.

The paper is organized as follows. We introduce the concept of rigidity and concept of a pinned point starting from the simplest case of 1-D motion in Sec.~\ref{sec:1D}, before progressing, in Sec.~\ref{sec:2body}, to their definitions and implications in 2-body systems in 3-D, and further to their implications for $N$-body quantum systems in Sec.~\ref{sec:Nbody}. We discuss, briefly, nonadditive potentials, multiparticle rigidity and multiparticle cusp conditions in Sec.~\ref{sec:disc}, before concluding with outlook in Sec.~\ref{sec:conc}. 

\section{The concepts of rigidity and pinned point in 1-D}
\label{sec:1D}

The 1-D Schr\"odinger equation for a particle of mass $m$ moving in a potential $v(x)$ takes the form of 
\begin{subequations}
\begin{equation}
H\psi_{\epsilon}(x) = \left[-\frac{\hbar^2}{2m}\frac{d^2}{dx^2} 
	+ v(x) \right]
	\psi_{\epsilon}(x) = \epsilon \psi_{\epsilon}(x),
\label{eq:sch1Da}
\end{equation}
or equivalently
\begin{equation}
\frac{d^2}{dx^2} \psi_{\epsilon}(x)
	+ \frac{2m}{\hbar^2}[\epsilon- v(x) ]
	\psi_{\epsilon}(x) = 0 \;,
\label{eq:sch1Db}
\end{equation}
\label{eq:sch1D}
\end{subequations}
where $\epsilon$ is the energy. We are interested in cases where $H$ is Hermitian, corresponding to $v(x)$ and $\epsilon$ being real. And we focus on real, ``standing wave'', solutions, which always exist in such cases \cite{LandauQM}, and from which other solutions, such as traveling wave solutions and scattering solutions, can be constructed as needed (see, e.g., \cite{Gao08a,Gao20b}). We introduce a concept of \textit{rigidity} to describe the sensitivity of a (real) wave function at a point in space to changes in energy, potential, and/or external perturbations, with the following considerations. 

The 1-D Schr\"odinger equation is a second-order homogeneous equation. Before imposing any boundary condition, there are two linearly independent solutions at each energy. Upon imposing a \textit{single} boundary condition on either $\psi_\epsilon|_{x=x_0}$, $\psi'_\epsilon|_{x=x_0}$, or $\psi_{\epsilon}'(\psi_{\epsilon})^{-1}|_{x=x_0}$ at $x_0$ (here prime refers to the spatial derivative), we have \textit{one solution at every energy $\epsilon$, determined up to a normalization constant}. For each such solution $\psi_\epsilon$, there is a corresponding log-derivative defined by
\begin{equation}
\mathcal{L}(x,\epsilon) := \psi_{\epsilon}'(x)\left[\psi_{\epsilon}(x)\right]^{-1} \;,
\end{equation}
and a $R$-matrix (see, e.g., Refs.~\cite{Aymar96,Burke11}) defined by
\begin{equation}
\mathcal{R}(x,\epsilon) := \psi_{\epsilon}(x)\left[\psi_{\epsilon}'(x)\right]^{-1} \;,
\end{equation}
which are the inverse of each other. Either quantity, together with the initial condition, has all the information contained in $\psi_\epsilon$. Unlike $\psi_\epsilon$, however, $\mathcal{L}$ and $\mathcal{R}$ are unique, independent of the normalization constant which is a global property.

In defining the concept of rigidity, the specific wave functions that we are referring to are the 1-boundary-condition (1BC) solutions $\psi_\epsilon$ mentioned above. They exist at \textit{all} energies and are determined up to a normalization constant. They each correspond to a unique log-derivative function $\mathcal{L}(x,\epsilon)$ and a unique $R$-matrix function $\mathcal{R}(x,\epsilon)$.

The simplest measure of the energy dependence of a function is its energy derivative. Since we would like our concept of rigidity to measure the energy dependence locally, we would like it to be related to either 
\[
\frac{\partial\mathcal{L}}{\partial\epsilon} = \left(\frac{\partial\psi_\epsilon'}{\partial\epsilon}
	\right)\left(\psi_\epsilon\right)^{-1}
	- \psi_\epsilon' \left(\frac{\partial\psi_\epsilon}{\partial\epsilon}\right)
	\left(\psi_\epsilon\right)^{-2} \;,
\]
or
\[
\frac{\partial\mathcal{R}}{\partial\epsilon} = \frac{\partial\psi_\epsilon}{\partial\epsilon}
	\left(\psi_\epsilon'\right)^{-1}
	- \psi_\epsilon
	\left(\frac{\partial\psi_\epsilon'}{\partial\epsilon}\right)		
	\left(\psi_\epsilon'\right)^{-2} \;,
\]
both locally unique and independent of the global normalization constant. Either quantity, however, would have limitation without adjustment. This is in addition to the decision as to which of the two to choose. The $\partial\mathcal{R}/\partial\epsilon$ is singular at ordinary nodes of $\psi_{\epsilon}'(x)$. Such singularities cannot be physically meaningful in the definition of rigidity as the corresponding $\partial\mathcal{L}/\partial\epsilon$ are finite at the same locations. The simple derivative $\partial\mathcal{L}/\partial\epsilon$ has similar difficulties at the ordinary nodes of $\psi_{\epsilon}(x)$. Fortunately all issues can be resolved simultaneously by recognizing that both quantities, properly regularized, become the same. More specifically,
\[
\left(\frac{\partial\mathcal{R}}{\partial\epsilon}\right) \left(\psi_\epsilon'\right)^2
	= - \left(\frac{\partial\mathcal{L}}{\partial\epsilon}\right) \left(\psi_\epsilon\right)^2 \;,
\]
by definitions of $\mathcal{L}$ and $\mathcal{R}$, which suggests that there exists a unique quantity that is a good measure of the rigidity of a wave solution.

We define the rigidity of a wave function, specifically of the 1BC solution $\psi_{\epsilon}$, as  
\begin{subequations}
\begin{align}
\mathscr{G}(x,\epsilon) &:= - \left[\frac{\hbar^2}{2m}\left(\frac{\partial\mathcal{L}}{\partial\epsilon}\right) 
	\left(\psi_\epsilon\right)^2\right]^{-1} \;,\\
	&= \left[\frac{\hbar^2}{2m}
	\left(\frac{\partial\mathcal{R}}{\partial\epsilon}\right) \left(\psi_\epsilon'\right)^2\right]^{-1} \;.
\end{align}
\label{eq:rgdef1D}
\end{subequations}
The symbol $\mathscr{G}$ is taken from ri$\mathscr{G}$idity, considering that the first letter $R$ has already many other widely used meanings such as the $R$-matrix and the hyperradius. Thus the rigidity is basically a derivative of the $R$-matrix with respect to energy or a derivative of the log-derivative with respect to energy, regularized to rid of the artificial singularities at either the regular nodes of $\psi_\epsilon'$ or the regular nodes of $\psi_\epsilon$. The inverse is taken such that ``more rigid'', namely greater rigidity $\mathscr{G}$, should corresponds to a smaller $|\partial\mathcal{L}/\partial\epsilon|$ or $|\partial\mathcal{R}/\partial\epsilon|$, namely being less sensitive to energy. The constant in the definition is chosen such that $\mathscr{G}$ is a dimensionless number when $\psi_\epsilon$ is normalized to 1. This concept of rigidity is well defined at every energy, corresponding to every 1BC solution $\psi_\epsilon$. Before discussions of its properties, we first note, from Eq.~(\ref{eq:sch1Db}), that $v\to v+\Delta v$, representing a change in potential or the application of an external perturbation, is equivalent to $\epsilon\to \epsilon-\Delta v$. Thus a local measurement of the sensitivity to energy is simultaneously a local measurement of the sensitivity to a change in potential or to an external perturbation.

The properties of rigidity are closely tied to the concept of a pinned point, which is a generalization of a similar concept in the description of a vibrating string. Specifically, we define a pinned point, $x_p$, to be a point where the wave function satisfies the boundary condition of $\psi_\epsilon|_{x=x_p}=0$ \textit{independent of energies}. It can be understood as a \textit{special node} of the wave function $\psi_\epsilon(x)$ that \textit{does not move with energy}. For comparison, an ordinary node would move with energy.

Let there be a pinned point $x_p$ in 1-D motion. For easy comparison and generalization to 3-D, we can take this point, without any loss of generality, to be the origin, $x_p=0$, and focuses on the characteristics of the resulting 1BC solution, $\psi_\epsilon$, in the region of $x>0$. In other words, we focus on the special set of 1BC solutions that are pinned at the origin, satisfying the boundary condition of $\psi_\epsilon|_{x=0}=0$ independent of energy. For this set of solutions, one can show from the Schr\"odinger equation that there exists a fundamental relation between the probability of finding the particle within $(0, x)$ around the pinned point, and the energy derivative of either the log-derivative or the $R$-matrix of the wave function, at the boundary point $x$. Specifically,
\begin{subequations}
\begin{align}
\int_0^x [\psi_{\epsilon}(x')]^2 dx' &= -\frac{\hbar^2}{2m}
	\left(\frac{\partial\mathcal{L}(x,\epsilon)}{\partial\epsilon}\right)
	\left[\psi_\epsilon(x)\right]^2 \;,\\
	&= \frac{\hbar^2}{2m}\left(\frac{\partial\mathcal{R}(x,\epsilon)}{\partial\epsilon}\right)
	\left[\psi'_\epsilon(x)\right]^2 \;.
\end{align}
\label{eq:rgR1D} 
\end{subequations}
This relation, in its 3-D form for the log-derivative, was given years ago in a nearest-neighbor theory for a Bose gas or liquid \cite{Gao05b}. An outline of the derivation is given in Appendix~\ref{sec:rgFRderiv}. In terms of rigidity, as defined by Eq.~(\ref{eq:rgdef1D}), Eq.~(\ref{eq:rgR1D}) can be rewritten as
\begin{equation}
\mathscr{G}(x,\epsilon) = \left[\int_0^x [\psi_{\epsilon}(x')]^2 dx'\right]^{-1} \;,
\label{eq:rgFR1D} 
\end{equation}
which we call the fundamental result for rigidity. 

The mathematical relation, Eq.~(\ref{eq:rgR1D}), was originally developed for the purpose of the evaluation of the normalization type of integrals on its left-hand side \cite{Gao05b}. The introduction of the rigidity concept helps to reveal its further and deeper implications. We summarize the key points as follows, all of which are easily observed or derived from Eqs.~(\ref{eq:rgFR1D}) and  (\ref{eq:rgdef1D}).
\begin{enumerate}[i.]
\item The rigidity $\mathscr{G}$ is everywhere positive for $x>0$, and it is a monotonically decreasing function of $x$, namely in the direction away from the pinned point. This is obvious from Eq.~(\ref{eq:rgFR1D}).

\item The property of $\mathscr{G}>0$ everywhere for $x>0$ implies that 
\begin{equation}
\frac{\partial\mathcal{L}}{\partial\epsilon} = -\frac{2m}{\hbar^2}\mathscr{G}^{-1}\left(\psi_\epsilon\right)^{-2} < 0 \;.
\end{equation}
It means that the log-derivative of a 1BC solution, $\psi_\epsilon$, pinned at the origin, is a piecewise monotonically decreasing function of energy at any point $x>0$. Similarly we have
\begin{equation}
\frac{\partial\mathcal{R}}{\partial\epsilon} = \frac{2m}{\hbar^2}\mathscr{G}^{-1}\left(\psi_\epsilon'\right)^{-2} > 0 \;.
\end{equation}
It means that the $R$-matrix of a 1BC solution, $\psi_\epsilon$, pinned at the origin, is a piecewise monotonically increasing function of energy, at any point $x>0$.

\item A pinned point is a singular point of the rigidity, as reflected in $\mathscr{G}\overset{x\to 0}{\sim} \infty$ from Eq.~(\ref{eq:rgFR1D}). Mathematically and physically, its implication can be further understood from
\[
\frac{\partial \mathcal{L}}{\partial\epsilon} \overset{x\to 0}{\sim} 
	O(x) \overset{x\to 0}\sim 0 \;,
\]
and
\[
\frac{\partial \mathcal{R}}{\partial\epsilon} \overset{x\to 0}{\sim} 
	O(x^3) \overset{x\to 0}\sim 0 \;.
\]
Namely, both the log-derivative and the $R$-matrix are independent of energy in the limit of $x\to 0$. For the wave function itself, it means that it can depend on energy only through a normalization constant in the limit of $x\to 0$, 
\begin{equation}
\psi_{\epsilon}(x) \overset{x\to 0}{\sim} c^{cp}_{\epsilon} F^{cp}(x)\;,
\label{eq:cuspcdef1D}
\end{equation}
where $F^{cp}(x)$ is a function of $x$ that is independent of energy. 

\item Moving away from the pinned point where the rigidity is infinite, a 1BC solution $\psi_\epsilon$ remains largely rigid, namely remains nearly energy-independent aside from the normalization constant, throughout the region of small probability, until a region is reached where the particle can be found with a substantial probability.

\end{enumerate}
The property~iii, specifically Eq.~(\ref{eq:cuspcdef1D}), defines what we call the \textit{strict cusp condition} in 1-D that comes with every pinned point. We call the energy-dependent normalization constant in Eq.~(\ref{eq:cuspcdef1D}), $c^{cp}_{\epsilon}$, the \textit{cusp amplitude}. And we call the energy-independent function, $F^{cp}(x)$, the \textit{strict cusp function}. The details of the strict cusp functions are determined by the behavior of the potential in the neighborhood of the pinned point, namely by $\lim_{x\to 0} v(x)$, and can be classified accordingly. We defer further developments and discussions to the next section since 1-D cusp conditions are contained as a special case of 2-particle cusp conditions in 3-D, corresponding to that of an $s$ wave ($\ell=0$).

There is another reason that we do not dwell on the 1-D case much further. In 1-D motion, a pinned point, thus the cusp condition associated with it, does not necessarily have to exist. It needs to be imposed. Specifically, a 1-D wave function needs to be pinned either by a singular potential, such as a hard wall at $x=0$, or by symmetry, specifically the parity conservation, $v(-x)=v(x)$, which would pin the odd-parity states at $x=0$. While it is important that the concept of rigidity and the concept of a pinned point are well defined and useful in 1-D motion, what makes them fundamental is their omnipresence in $N$-body quantum systems with $N\ge 2$ in 3-D, to be discussed in the next two sections.

\section{2-particle rigidity and its implications for 2-body systems in 3-D}
\label{sec:2body}

We introduce here the concept of rigidity for 2-body systems in 3-D, and derive and discuss in detail its substantial set of consequences. They include definitions of two related cusp functions that we call the \textit{cusp function} and the \textit{strict cusp function}, the general analytic structure of a 2-body regular solution versus energy, and its relation to the cusp functions. We further reveal universal behaviors in cusp conditions and provide a classification of quantum system based on them, using a method that can be called the QDT for cusp functions, or QDT for short-range correlation, to emphasize both its similarities and its differences from the more standard QDTs \cite{Seaton83,Greene79,Greene82,Greene85,Gao08a}, which have traditionally been about the long range behavior only.

\subsection{2-body rigidity}

While a pinned point needs to be imposed in 1-D, what makes the concept universally important is that it arises naturally in any $N$-body quantum system with $N\ge 2$, regardless of the interaction potential (including the case of no interaction), as a part of the physical constraints on the wave function \cite{Dirac58,LandauQM}. 

For a 2-body system interacting via a central potential $v(r)$, the relative motion in the center-of-mass frame is described by (see, e.g., \cite{LandauQM})
\begin{equation}
\left[-\frac{\hbar^2}{2\mu}\nabla_r^2
	+ v(r) \right]
	\psi_{\epsilon} = \epsilon \psi_{\epsilon}.
\label{eq:sch3D}
\end{equation}
Its solution can always be written as
\[
\psi_{\epsilon} = \sum_{\ell,m}a_{\ell m}\frac{u_{\epsilon\ell}(r)}{r}Y_{\ell,m}(\theta,\phi) \;,
\]
where $Y_{\ell,m}$ is the standard spherical harmonics \cite{Olver10,Arfken13}, $a_{\ell m}$'s are the constants of superposition. The radial wave function, $u_{\epsilon\ell}(r)$, satisfies an effective 1-D Schr\"odinger equation, defined in $0\le r<\infty$ for a reduced mass $\mu$,
\begin{equation}
\left[-\frac{\hbar^2}{2\mu}\frac{d^2}{dr^2} 
	+ \frac{\hbar^2 \ell(\ell+1)}{2\mu r^2}
	+ v(r) \right]
	u_{\epsilon\ell}(r) = \epsilon u_{\epsilon\ell}(r),
\label{eq:rsch}
\end{equation}
and what we call the physical boundary condition of 
\begin{subequations}
\begin{equation}
u_{\epsilon\ell}|_{r=0} = 0 \;,
\end{equation}
or more generally,
\begin{equation}
u_{\epsilon\ell} \overset{r\to 0}{\sim} 0 \;.
\end{equation}
\label{eq:PBC3D}
\end{subequations}
This boundary condition, satisfied by \textit{all} physical solutions of a 2-body system, is due to a combination of the physical interpretation of a wave function and the mathematical self-consistency of the theory \cite{Dirac58,LandauQM}. Compared to our discussions of 1-D equation and 1-D pinned point, it clear that for a 2-body quantum system interacting via a central potential, the origin, more precisely the point of 2-particle coalescence, is a naturally pinned point for the two-body relative radial motion. It is naturally pinned in the sense that it needs no external condition, such as an infinite potential, to impose. 

With this understanding, all results of Sec.~\ref{sec:1D} can be generalized to the 3-D case for each partial wave $\ell$, with the substitutions of $m$, $x$, and $\psi_\epsilon$ by $\mu$, $r$, and $u_{\epsilon\ell}$, respectively. For simplicity and consistent with standard convention, we call the 1BC physical solution satisfying the boundary condition of Eq.~(\ref{eq:PBC3D}), $u_{\epsilon\ell}$, the regular solution for partial wave $\ell$. In terms of this regular solution, the $R$-matrix (see, e.g., Refs.~\cite{Aymar96,Burke11}) for partial wave $\ell$ is defined by
\begin{equation}
\mathcal{R}_\ell(r,\epsilon) := u_{\epsilon\ell}(r)\left[u_{\epsilon\ell}'(r)\right]^{-1} \;.
\end{equation}
The log-derivative is defined by
\begin{equation}
\mathcal{L}_{\ell}(r,\epsilon) := u_{\epsilon\ell}'(r)\left[u_{\epsilon\ell}(r)\right]^{-1} \;.
\end{equation}
There is one unique $\mathcal{R}_\ell$ and one unique $\mathcal{L}_\ell$ for each partial wave and for every energy.

For each regular solution $u_{\epsilon\ell}$, we define the rigidity for partial wave $\ell$ as [cf. Eq.~(\ref{eq:rgdef1D})]
\begin{subequations}
\begin{align}
\mathscr{G}_\ell(r,\epsilon) &= \left[\frac{\hbar^2}{2\mu}
	\left(\frac{\partial\mathcal{R}_\ell}{\partial\epsilon}\right) \left(u_{\epsilon\ell}'\right)^2\right]^{-1} \;, \\
	&= - \left[\frac{\hbar^2}{2\mu}\left(\frac{\partial \mathcal{L}_{\ell}}{\partial\epsilon}\right) 
	\left(u_{\epsilon\ell}\right)^2\right]^{-1} \;.
\end{align}
\label{eq:rgdef3D}
\end{subequations}
It is well defined for every partial wave and every energy, including negative energies and energies below the minimum of the potential.

Similar to the 1-D case, we have (see \cite{Gao05b} and Appendix~\ref{sec:rgFRderiv})
\begin{subequations}
\begin{align}
\int_0^r [u_{\epsilon\ell}(r')]^2 dr' &= -\frac{\hbar^2}{2\mu}
	\left(\frac{\partial\mathcal{L}_{\ell}(r,\epsilon)}{\partial\epsilon}\right)
	\left[u_{\epsilon\ell}(r)\right]^2 \;,\\
	&= \frac{\hbar^2}{2\mu}
	\left(\frac{\partial\mathcal{R}_{\ell}(r,\epsilon)}{\partial\epsilon}\right)
	\left[u'_{\epsilon\ell}(r)\right]^2 \;.
\end{align}
\label{eq:rgR3D}
\end{subequations}
It gives the fundamental result for rigidity in 3-D,
\begin{equation}
\mathscr{G}_\ell(r,\epsilon) = \left[\int_0^r [u_{\epsilon\ell}(r')]^2 dr'\right]^{-1}\;,
\label{eq:rgFR3D}
\end{equation}
which relates rigidity to the probability of finding the pair to be within $(0,r)$. Contained in this relation and in the definition of rigidity are the following general properties of \textit{any} regular solution for 2-body relative motion in 3-D. They are sufficiently important to repeat here despite their similarities to the 1-D case.
\begin{enumerate}
\item For any regular solution $u_{\epsilon\ell}(r)$ describing 2-body relative motion in partial wave $\ell$, the rigidity $\mathscr{G}_{\ell}$ is always positive, and it is a monotonically decreasing function of $r$, namely in the direction away from the pinned point.

\item The property of $\mathscr{G}_\ell>0$ everywhere implies that 
\begin{equation}
\frac{\partial \mathcal{L}_{\ell}}{\partial\epsilon} = -\frac{2\mu}{\hbar^2}\mathscr{G}_\ell^{-1}\left(u_{\epsilon\ell}\right)^{-2} < 0 \;.
\end{equation}
It means that for any regular solution $u_{\epsilon\ell}(r)$ describing 2-body relative motion, the log-derivative at any point in space is a piecewise monotonically decreasing function of energy. Similarly we have
\begin{equation}
\frac{\partial\mathcal{R}_\ell}{\partial\epsilon} = \frac{2\mu}{\hbar^2}\mathscr{G}_\ell^{-1}\left(u_{\epsilon\ell}'\right)^{-2} > 0 \;.
\end{equation}
It means that for any regular solution $u_{\epsilon\ell}(r)$ describing 2-body relative motion, the $R$-matrix at any point in space is a piecewise monotonically increasing function of energy.

\item The 2-body coalescence point, $r=0$, is a naturally pinned point, where the rigidity is infinite, independent of the pair potential and including the case of free particles. Mathematically and physically, this is further understood as
\begin{equation}
\frac{\partial \mathcal{L}_{\ell}}{\partial\epsilon} \overset{r\to 0}{\sim} 
	O(r) \overset{r\to 0}\sim 0 \;,
\label{eq:LDr0}
\end{equation}
and
\begin{equation}
\frac{\partial \mathcal{R}_{\ell}}{\partial\epsilon} \overset{r\to 0}{\sim} 
	O(r^3) \overset{r\to 0}\sim 0 \;.
\label{eq:RMr0}
\end{equation}
Namely, both the log-derivative and the $R$-matrix are independent of energy in the limit of $r\to 0$. For the wave function itself, it means that it can depend on energy only through a normalization constant in the limit of $r\to 0$,
\begin{equation}
u_{\epsilon\ell} \overset{r\to 0}{\sim} c^{cp}_{\epsilon\ell} F^{cp}_{\ell}(r)\;,
\label{eq:stcuspcdef}
\end{equation}
where $F^{cp}_{\ell}(r)$ is a function of $r$ that is independent of energy. 

\item Moving away from $r=0$ where the rigidity is infinite, the regular solution $u_{\epsilon\ell}$ remains largely rigid, namely remains nearly energy-independent aside from the normalization constant, throughout the region of small probabilities, until a region is reached where the particle can be found with a substantial probability.

\end{enumerate}

The Property~1 is a general property of rigidity that leads to Property~2 and in part to Property~4. The Property~2, specifically the $R$ matrix being a piecewise monotonically increasing function of energy, has also been derived in the $R$-matrix theory by Wigner and Eisenbud \cite{Wigner47}. The Properties~3 and 4, and some of their consequences, are the further focus of this section.
The Property~3, specifically Eq.~(\ref{eq:stcuspcdef}), defines what we call the \textit{strict cusp condition} for 2-particle relative motion in 3-D. We call the energy-dependent normalization constant in Eq.~(\ref{eq:stcuspcdef}), $c^{cp}_{\epsilon\ell}$, the \textit{cusp amplitude}. And we call the energy-independent function, $F^{cp}_{\ell}(r)$, the \textit{strict cusp function}. The details of the strict cusp function depend on the behavior of the potential in the neighborhood of the pinned point, namely by $\lim_{r\to 0} v(r)$, and can be classified accordingly. This will be done in the next section. We note here that contained in the strict cusp function are the energy-independent limits for the log-derivative and the $R$-matrix, as implied by Eqs.~(\ref{eq:LDr0}) and (\ref{eq:RMr0}), and given explicitly in terms of the $F^{cp}_{\ell}(r)$ by
\begin{equation}
\mathcal{L}_{\ell} \overset{r\to 0}{\sim} 
	\frac{d F^{cp}_\ell(r)}{dr}\left[F^{cp}_\ell(r)\right]^{-1}\;,
\label{eq:LDstcuspC}
\end{equation}
\begin{equation}
\mathcal{R}_{\ell} \overset{r\to 0}{\sim} 
	F^{cp}_\ell(r)\left[\frac{d F^{cp}_\ell(r)}{dr}\right]^{-1} \;.
\label{eq:RMstcuspC}
\end{equation}
Thus the essence of a cusp condition, stated in simplest terms as having an energy-independent log-derivative or $R$-matrix at a 2-particle coalescence, is not a result of the singularity of a potential, as has been generally believed \cite{Kato57,Pack66,Gruneis17}. It is a general property at a 2-particle coalescence, attributable to the physical boundary condition of Eq.~(\ref{eq:PBC3D}) that makes the coalescence point a pinned point. We call all mutually related limiting behaviors at a 2-particle coalescence, Eqs.~(\ref{eq:stcuspcdef}), (\ref{eq:LDstcuspC}), and (\ref{eq:RMstcuspC}), \textit{strict cusp conditions}. 

\subsection{Analytic structure of a regular solution versus energy}

The Property~4 associated with the rigidity means that the energy dependence of a regular solution, excluding that of the normalization constant, can only develop gradually away from the coalescence point. One of its implications is the possibility of expanding the strict cusp conditions of Eqs.~(\ref{eq:stcuspcdef})-(\ref{eq:RMstcuspC}) to cover a greater range of $r$, if a limit on the energy range of interest is imposed. In other words, there should exist an energy-independent $f^{cp}_{\ell}(r)$, with the property of 
\begin{equation}
f^{cp}_{\ell}(r) \overset{r\to 0}{\sim} F^{cp}_{\ell}(r) \;,
\label{eq:fcpFcp}
\end{equation}
such that
\begin{equation}
u_{\epsilon\ell} \sim c^{cp}_{\epsilon\ell} f^{cp}_{\ell}(r)\;,
\label{eq:cuspcdef}
\end{equation}
is satisfied beyond the limit of $r\to 0$, provided a constraint is put on the range of $\epsilon$. This expansion of the strict cusp conditions, exact in the limit of strict cusp limit of $r\to 0$, to cusp conditions representing asymptotic behaviors applicable over a greater range of $r$,  can be done in way that further reveals the general analytic structure of the regular solution $u_{\epsilon\ell}(r)$ versus energy.

Corresponding to the strict form of the cusp condition, Eq.~(\ref{eq:stcuspcdef}), one can always define a special regular solution, $u^{cp}_{\epsilon\ell}$, which we call the \textit{cusp solution}, through
\begin{subequations}
\begin{equation}
u_{\epsilon\ell} = c^{cp}_{\epsilon\ell}u^{cp}_{\epsilon\ell} \;,
\label{eq:ucpdef1}
\end{equation}
with 
\begin{equation}
u^{cp}_{\epsilon\ell}(r) \overset{r\to 0}{\sim} F^{cp}_{\ell}(r) \;.
\label{eq:ucpdef2}
\end{equation}
\label{eq:ucpdef}
\end{subequations}
The cusp solution, $u^{cp}_{\epsilon\ell}$, is thus a special regular solution with an energy-independent behavior near the point of coalescence as specified by the strict cusp function $F^{cp}_{\ell}(r)$.  Such a regular solution, with its energy-independent initial conditions for both $u^{cp}_{\epsilon\ell}$ and $du^{cp}_{\epsilon\ell}/dr = d F^{cp}_{\ell}/dr$, 
is an \textit{entire function of energy} at any finite $r$, with only a minor requirement of $v(r')$ being continuous up to $r$, namely in the region $0<r'<r$ \cite{Poincare1884,JP51,Lin88}. Such a function can always be written, using its Taylor expansion in energy, as
\begin{subequations}
\begin{equation}
u^{cp}_{\epsilon\ell}(r) = f^{cp}_{\ell}(r) 
	+ \sum_{j=1}^\infty \epsilon^j x^{(j)}_{\ell}(r) \;,
\label{eq:ucpexpa}
\end{equation}
with
\begin{equation}
f^{cp}_{\ell}(r) = u^{cp}_{\epsilon=0,\ell}(r) \;,
\label{eq:fcpl}
\end{equation}
and
\begin{equation}
x^{(j)}_{\ell}(r) = \frac{1}{j!}\left.
	\frac{\partial^j u^{cp}_{\epsilon\ell}(r)}{\partial\epsilon^j}\right|_{\epsilon=0} \;.
\end{equation}
\label{eq:ucpexp}
\end{subequations}
Any regular solution, because it has to satisfy the strict cusp condition of Eq.~(\ref{eq:stcuspcdef}), can differ from the cusp solution only by a normalization constant, as is already implied by Eq.~(\ref{eq:ucpdef1}). Equations~(\ref{eq:ucpdef}) and (\ref{eq:ucpexp}) thus provide the general analytic structure as a function of energy followed by \textit{all} regular solutions for all potentials that are continuous in $(0,\infty)$. It is sufficiently important that we regroup the result as
\begin{align}
u_{\epsilon\ell} &= c^{cp}_{\epsilon\ell}u^{cp}_{\epsilon\ell} \;,\nonumber\\
	&= c^{cp}_{\epsilon\ell}
	\left[ f^{cp}_{\ell}(r) 
	+ \sum_{j=1}^\infty \epsilon^j x^{(j)}_{\ell}(r) \right] \;,
\label{eq:ue}
\end{align}
where $f^{cp}_{\ell}(r)$, and $x^{(j)}_{\ell}(r)$ for all $j$, are all \textit{independent of energy}.  The condition of Eq.~(\ref{eq:ucpdef2}), which has to be satisfied for all energies as a result of $r=0$ being a point of infinite rigidity, further implies that $x^{(i)}_{\ell}(r)$ has to satisfies
\begin{equation}
x^{(j)}_{\ell}(r)/f^{cp}_{\ell}(r) \overset{r\to 0}{\sim} x^{(j)}_{\ell}(r)/F^{cp}_{\ell}(r) \overset{r\to 0}{\sim} 0 \;,
\label{eq:xlr0}
\end{equation}
for all $j=1,2,\cdots$. Such a regular solution exists at all energies. 

One of the implications of this analytic structure versus energy is that the strict cusp condition of Eq.~(\ref{eq:stcuspcdef}) is now generalized to Eq.~(\ref{eq:cuspcdef}), which we call the \textit{cusp condition}, applicable over a broader range of $r$. This range can approach infinite for $\epsilon$ within a sufficiently small range around $\epsilon=0$. More precisely, at every energy, Eq.~(\ref{eq:cuspcdef}) is satisfied for sufficiently small $r$, and at every finite $r$, it is satisfied for sufficiently small energy. We group the conditions together to write
\begin{equation}
u_{\epsilon\ell} \sim c^{cp}_{\epsilon\ell} f^{cp}_{\ell}(r)
	\overset{r\to 0}{\sim} c^{cp}_{\epsilon\ell} F^{cp}_{\ell}(r)\;,
\label{eq:uelcuspC}
\end{equation}
which is a more complete statement of cusp conditions. We call $f^{cp}_{\ell}(r)$ the \textit{cusp function} to distinguish it from the strict cusp function $F^{cp}_{\ell}(r)$. Embedded in the same more complete statement of cusp conditions is a further understanding that the two limits of the cusp solution, the strict cusp limit of $r\to 0$ described by $F^{cp}_{\ell}(r)$ as in Eq.~(\ref{eq:ucpdef2}), and the limit of $\epsilon\to 0$ described by $f^{cp}_{\ell}(r)$, 
\begin{equation}
u^{cp}_{\epsilon\ell}(r) \overset{\epsilon\to 0}{\sim} f^{cp}_{\ell}(r) \;,
\label{eq:ucpfcp}
\end{equation}
as implied by Eq.~(\ref{eq:ucpexpa}), are related, by Eq.~(\ref{eq:fcpFcp}), (but generally different except for potentials that do not have a length scale, as will become evident later in the paper).

Another implication of $u_{\epsilon\ell}$, up to a normalization constant, being an entire function of energy is that the corresponding $\mathcal{R}_\ell$ and $\mathcal{L}_\ell$ are meromorphic functions of energies, a conclusion that is also known from the $R$-matrix theory \cite{Wigner47,Burke11}. In terms of the entire cusp solution $u^{cp}_{\epsilon\ell}$, they can be written explicitly as
\begin{equation}
\mathcal{L}_{\ell} = 
	\frac{d u^{cp}_{\epsilon\ell}(r)}{dr}\left[u^{cp}_{\epsilon\ell}(r)\right]^{-1}\;,
\label{eq:LDucp}
\end{equation}
\begin{equation}
\mathcal{R}_{\ell} = 
	u^{cp}_{\epsilon\ell}(r)\left[\frac{d u^{cp}_{\epsilon\ell}(r)}{dr}\right]^{-1} \;,
\label{eq:RMucp}
\end{equation}
leading to the generalizations of their strict cusp conditions of Eqs.~(\ref{eq:LDstcuspC}) and (\ref{eq:RMstcuspC}) to more complete cusp conditions
\begin{equation}
\mathcal{L}_{\ell} {\sim} 
	\frac{d f^{cp}_\ell(r)}{dr}\left[f^{cp}_\ell(r)\right]^{-1}
	\overset{r\to 0}{\sim} 
	\frac{d F^{cp}_\ell(r)}{dr}\left[F^{cp}_\ell(r)\right]^{-1}\;,
\label{eq:LDcuspC}
\end{equation}
\begin{equation}
\mathcal{R}_{\ell} {\sim} 
	f^{cp}_\ell(r)\left[\frac{d f^{cp}_\ell(r)}{dr}\right]^{-1}
	\overset{r\to 0}{\sim} 
	F^{cp}_\ell(r)\left[\frac{d F^{cp}_\ell(r)}{dr}\right]^{-1} \;,
\label{eq:RMcuspC}
\end{equation}
with meanings similar to those discussed earlier for $u_{\epsilon\ell}$.

The analytic structure of a regular solution versus energy, its relations to the cusp conditions, and the constraints specified by Eq.~(\ref{eq:xlr0}), which are all derived here from the concept of rigidity, are some of the key results of this work. They are fundamental results on wave function from which many other results, including the $N$-body cusp conditions of Sec.~\ref{sec:Nbody}, are derived.

Embedded in our derivation and discussion is also a practical method to find the cusp functions. Specifically, the cusp function $f^{cp}_{\ell}(r)$ is obtained, using Eq.~(\ref{eq:fcpl}),  as the zero-energy cusp solution of the radial equation, namely a regular solution of the radial equation at the zero energy with an energy-independent normalization near the origin. From $f^{cp}_{\ell}(r)$, the strict cusp function can be obtained from [cf. Eq.~(\ref{eq:fcpFcp})]
\begin{equation}
F^{cp}_{\ell}(r) = \lim_{r\to 0}f^{cp}_{\ell}(r) \;.
\label{eq:stCuspFcusp} 
\end{equation}
This method, which highlights the special role of the zero-energy solution both in the analytic structure versus energy and in cusp conditions, will be amply illustrated in later examples.

Our analysis greatly expands the existing understanding of the analytic properties of a wave function versus energy (see, e.g., Refs.~\cite{JP51,Newton60,Joachain75,Newton82,Taylor06}), not only in the extra constraints and in their relations to cusp conditions, but also in the scope of potentials we cover. The previous studies have focused on potentials satisfying the Jost-Pais-Newton (JPN) criterion \cite{JP51,Newton60} of
\begin{equation}
\int_{0}^\infty dr r|v(r)|<\infty \;,
\label{eq:JPc}
\end{equation}
or its equivalent which requires that the potential goes to zero faster than $1/r^2$ at infinity and diverge slower than $1/r^2$ at the origin. This latter requirement, as will become clear in Sec.~\ref{sec:2PcuspF}, would have limited us only to a class of potentials with $r^{\ell+1}$ type of strict cusp conditions for the wave function. Our analysis is much broader in scope and covers all potentials that are continuous in $(0,\infty)$, which, we believe, includes all ``real'' physical potentials. The broader scope in potential will be reflected in more classes of cusp conditions, to be addressed in Sec.~\ref{sec:2PcuspF}. We also point out that the JPN criterion, due partly to its original focus on the analytic properties of scattering, even excludes some of the well-known potentials such as the linear (Stark) potential and the harmonic-oscillator potential. Our analysis has no such exclusion.

\subsubsection{An example of the free-particle (F) class}
\label{sec:Fclass}

The free-particle (F) class, corresponding to 
\[
v(r) = 0 \;,
\]
everywhere, provides a simple example with analytic solutions. It also provides a reference point with respect to which the effects of interaction can be understood.

The cusp function $f^{cp}_{\ell}(r)$ for the F class, to the denoted by $f^{cp(F)}_{\ell}(r)$, is easily found from its zero energy solution (see, e.g., \cite{LandauQM}). We obtain
\begin{equation}
f^{cp(F)}_{\ell}(r, s_{L}) =  
	\frac{1}{2^{\ell+\tfrac{1}{2}}\Gamma(\ell+\tfrac{3}{2})} (r/s_{L})^{\ell+1}\;.
\label{eq:fcpF}
\end{equation}
Here $s_L$ is a length scale that makes $r^{\ell+1}$ term dimensionless. Since the free-particle potential has no intrinsic length scale by itself, $s_L$ can be understood here as the unit one chooses for $r$, and as a part of the energy-independent normalization constant (see also Appendix~\ref{sec:fcp1tDeriv}). It is a ``free'' length scale that is ready to be replaced by other length scales whenever they arise in the system.

The strict cusp function, $F^{cp}_{\ell}(r)$ for the F class, to the denoted by $F^{cp(F)}_{\ell}(r)$, is obtained from $F^{cp(F)}_{\ell}(r) = \lim_{r\to 0}f^{cp(F)}_{\ell}(r)$, which, in the special case of the F class, is equal to $f^{cp(F)}_{\ell}(r)$ itself,
\begin{equation}
F^{cp(F)}_{\ell}(r, s_{L}) = f^{cp(F)}_{\ell}(r, s_{L}) = 
	\frac{1}{2^{\ell+\tfrac{1}{2}}\Gamma(\ell+\tfrac{3}{2})} (r/s_{L})^{\ell+1}\;.
\label{eq:FcpF}
\end{equation}
This equality means, from our earlier discussion, that for free particles the limits of $\epsilon\to 0$ and $r\to 0$ for the cusp solution are the same.

For the F class, we know the regular solution at all energies (see, e.g., \cite{LandauQM})
\[
u^{(F)}_{\epsilon\ell} = Cr j_\ell(kr) \;,
\]
where $j_{\ell}(x)$ is the spherical Bessel function \cite{Olver10,Arfken13}, $k:=\sqrt{2\mu\epsilon/\hbar^2}$ for all energies with the understanding of $k=i\kappa, \ \kappa:=\sqrt{2\mu(-\epsilon)/\hbar^2}$ for $\epsilon<0$. From the series expansion of the spherical Bessel function \cite{Olver10,Arfken13}, one can easily check that the solution can indeed by written in the form of Eq.~(\ref{eq:ue}), with $f^{cp}_\ell$ given by $f^{cp(F)}_{\ell}(r)$ and with
\[
x^{(j)}_\ell(r) = \frac{(-1)^j(2\mu/\hbar^2)^j}{2^{2j+\ell+1/2}j!\Gamma(j+\ell+3/2)}
	(r/s_L)^{\ell+1}r^{2j}\;.
\]
It is clear that the conditions of Eq.~(\ref{eq:xlr0}) are satisfied, and so are all the cusp conditions. For instance,
\begin{equation}
\mathcal{L}_{\ell}(r,\epsilon) \sim \frac{\ell+1}{r} \overset{r\to 0}{\sim} \frac{\ell+1}{r} \;,
\end{equation}
where the first asymptotic behavior is not only satisfied at sufficiently small $r$, but also at any finite $r$ for sufficiently small energies.

\subsection{Universal aspects of 2-particle cusp functions}
\label{sec:2PcuspF}

The cusp function $f^{cp}_{\ell}(r)$, up to an energy-independent normalization, is determined by the behavior of the potential $v(r')$ in the region from the origin up to $r$, namely in  $0<r'<r$. For sufficiently small $r$, $v(r)$ can be represented by the leading term(s) of its expansion around $r=0$, giving rise to, as we show here, a classification and a corresponding set of universal behaviors for both $f^{cp}_{\ell}(r)$ and $F^{cp}_{\ell}(r)$ according to the behavior of the potential around the coalescence.

We are interested in potentials $v(r)$ that are continuous in $0<r<\infty$, but can in general be singular with a pole of finite order at either $r=0$ or $r=\infty$. Such potentials can be broadly grouped into 2 categories according to its behavior at the origin: the regular classes for which $v(r)$ is analytic at $r=0$, with $\lim_{r\to 0} v(r)$ being a constant, and the singular classes for which $v(r)$ has a pole at $r=0$. 

A singular potential with a pole at $r=0$ can be written as $v(r) = D(r)/r^\alpha$ with $\alpha>0$ and $D(r)$ being analytic at $r=0$. It can be expanded around $r=0$ by expanding $D(r)$ in a power series.  The resulting expansion (the Laurent expansion of $v(r)$ around $r=0$ in the special case of $\alpha$ being an integer) can be written as
\[
v(r) =
\pm \frac{D_{\alpha}}{r^\alpha} +\sum_{\{\alpha'<\alpha\}} \frac{(\pm D_{\alpha'})}{r^{\alpha'}} \;,
\]
where we have chosen $D_{\alpha}>0$ and $D_{\alpha'}>0$ by definition and have isolated the leading term from the rest for easier discussion. The sum over $\alpha'$ is over successively smaller ones. If it does not terminates, it would become, after a finite number of terms, a sum over negative $\alpha'$ corresponding to a positive power of $r$, namely $\pm D_{\alpha'} r^{(-\alpha')}$ with $(-\alpha')>0$. A regular potential, regular at $r=0$, can be represented by formally the same expansion if we extend it to include the case of $\alpha\le 0$.

In all cases, the expansion of the potential around the origin is made of terms of the form of $\pm D_{\alpha}/r^\alpha$. Each such term has an important characteristic that for all $\alpha\neq 2$, including the cases of $\alpha=0$ and $\alpha<0$, the strength parameter $D_{\alpha}$ defines a length scale given by \cite{Gao08a}
\begin{equation}
\beta_{\alpha} = \left(2\mu D_{\alpha}/\hbar^2\right)^{1/(\alpha-2)} \;.
\label{eq:betadef}
\end{equation}
The case $\alpha=2$ is a special case that does not define a length scale. The $D_{2}$ parameter defines, instead, a \textit{dimensionless} strength parameter (see, e.g., \cite{LandauQM})  
\begin{equation}
\gamma_2 = 2\mu D_{2}/\hbar^2 \;.
\label{eq:gam2def}
\end{equation}

In the limit of $r\to 0$, all terms of $\alpha'<0$ go to zero, and we have the general asymptotic form for a singular potential
\begin{subequations}
\begin{equation}
v(r) \overset{r\to 0}{\sim} v^{(\{\alpha\})}(r) \;,
\end{equation}
where 
\begin{equation}
v^{(\{\alpha\})}(r) :=
\pm \frac{D_{\alpha}}{r^\alpha} +\sum_{\{\alpha'<\alpha\}}^{\alpha'>=0} \frac{(\pm D_{\alpha'})}{r^{\alpha'}} \;,
\end{equation}
\label{eq:vuo} 
\end{subequations}
consisting of, in general, a finite number of terms with the notation $\{\alpha\}$ representing the set of all exponents in the expansion. It is a Laurent polynomial if $\{\alpha\}$ are all integers, and can be called a generalized Laurent polynomial in other cases.

For a regular potential, all terms with $\alpha'<0$ go to zero. Keeping only the lowest-order term, we have
\[
v(r) \overset{r\to 0}{\sim} \pm \frac{D_{\alpha}}{r^\alpha} \;,
\]
with $\alpha\le 0$. This includes, as a special case, the case of $\alpha=0$, corresponding to $v(r)\sim v_0 = \pm D_0$. It is clear that the asymptotic behavior for a regular potential can be grouped into and regarded as a special, single-scale, case of the more general $v^{(\{\alpha\})}(r)$ of Eq.~(\ref{eq:vuo}).

This discussion shows that the behavior of a potential in the neighborhood of $r= 0$ can be generally represented by $v^{(\{\alpha\})}(r)$ of Eq.~(\ref{eq:vuo}), characterized by a set of exponents $\{\alpha\}$ and a corresponding set of strength parameters $\{D_{\alpha}\}$. A regular potential has an asymptotic behavior described by a single-term or a single-scale potential, characterized by a single $\alpha\le 0$. A singular potential may generally require a $v^{(\{\alpha\})}(r)$ with multiple $\alpha>0$, to be called a multi-term potential or a multiscale potential, in which case $r=0$ is called a multi-term singularity or multiscale singularity.

Since the cusp function $f^{cp}_{\ell}(r)$ at $r$ depends only on the potential in region $(0,r)$, and the strictly cusp function $F^{cp}_{\ell}(r)$ depends only on the asymptotic behavior of the potential near the origin, the classification of the asymptotic behavior of the potential near the origin implies a corresponding classification of $F^{cp}_{\ell}(r)$ and a corresponding set of universal behaviors for $f^{cp}_{\ell}(r)$, all based on the cusp solution for the $v^{(\{\alpha\})}(r)$ potential,
\begin{multline}
\left[-\frac{\hbar^2}{2\mu}\frac{d^2}{dr^2} 
	+ \frac{\hbar^2 \ell(\ell+1)}{2\mu r^2}
	+ v^{(\{\alpha\})}(r) \right]u^{(\{\alpha\})}_{\epsilon\ell}(r) \\
	= \epsilon u^{(\{\alpha\})}_{\epsilon\ell}(r).
\label{eq:rschuCusp1} 
\end{multline}
Specifically, its cusp solution at zero energy,
\begin{equation}
f^{cp(\{\alpha\})}_{\ell}(r) = u^{cp(\{\alpha\})}_{\epsilon=0,\ell}(r) \;,
\label{eq:uCusp2} 
\end{equation}
gives the cusp function for potential $v^{(\{\alpha\})}(r)$. Its corresponding strict cusp function is given by
\begin{equation}
F^{cp(\{\alpha\})}_{\ell}(r) = \lim_{r\to 0} f^{cp(\{\alpha\})}_{\ell}(r) \;.
\label{eq:uCusp3} 
\end{equation}
For any potential with an asymptotic behavior of Eq.~(\ref{eq:vuo}), its strict cusp function depends only on the asymptotic behavior and is given by the strict cusp function for $v^{(\{\alpha\})}(r)$, namely
\begin{equation}
F^{cp}_{\ell}(r) = \lim_{r\to 0} f^{cp}_{\ell}(r) = F^{cp(\{\alpha\})}_{\ell}(r) \;.
\end{equation}
And its cusp function follows the universal behavior of
\begin{equation}
f^{cp}_{\ell}(r) \overset{r<r_0}{\sim} f^{cp(\{\alpha\})}_{\ell}(r) \;, 
\label{eq:uCuspC} 
\end{equation}
where $r_0$ is the radius within which the potential $v(r)$ is well represented by $v^{(\{\alpha\})}(r)$ as in Eq.~(\ref{eq:vuo}). 

Thus potentials $v^{(\{\alpha\})}(r)$, similar in form to those of interest in QDTs \cite{Seaton83,Greene79,Greene82,Gao08a} and multiscale QDTs \cite{Gao16a,Hood20a}, play a special role in understanding cusp functions and cusp conditions. Their cusp functions not only give those for themselves, but also describe universal behaviors of all potentials with $v(r) \overset{r\to 0}{\sim} v^{(\{\alpha\})}(r)$. Understanding universal properties of cusp functions is in this sense mathematically similar to the theory of QDT reference functions \cite{Seaton83,Greene79,Greene82,Gao08a,Gao16a}. Both require solutions for potentials of the form of Eq.~(\ref{eq:vuo}). The main difference is in their focus. The cusp functions only require the regular solution at the zero energy. It is a far easier task than that in a full QDT formulation which would require pairs of linearly independent solutions for all energies \cite{Seaton83,Greene79,Greene82,Gao08a,Gao16a}. The use of the notation $D_{\alpha}$'s for the strength parameters in $v^{(\{\alpha\})}(r)$ helps to emphasize that they correspond to the expansion around the origin, and are generally different from (but can be same as) the parameters $C_{\alpha}$ (the van der Waals coefficients) describing the long-range potential \cite{Derevianko99,Stone13}. 

Similar to the treatment of long-range universal behavior in a QDT, the essence of universal behaviors of cusp functions is contained in the single-term (or single-scale) potential 
\begin{equation}
v^{(\alpha)}(r) := \pm \frac{D_{\alpha}}{r^\alpha} \;.
\label{eq:v1T} 
\end{equation}
It covers all regular potentials sufficiently close to the origin and most of the singular potentials of practical interest. The only exceptions, the intrinsically multi-term singular (ImtS) potentials that cannot be approximated by their dominant term, will be addressed when they arise in meaningful physical contexts in future studies, after related discussions in a general multiscale QDT (msQDT) for long-range behaviors where ImtS potentials are of greater interest. The relevant cusp functions and conditions in those case are still given by Eqs.~(\ref{eq:rschuCusp1})-(\ref{eq:uCuspC}). The difference is that $f^{cp(\{\alpha\})}_{\ell}(r)$ may no longer have an exact analytic solution, and will need a numerical or a semiclassical solution instead.

Through analytic solutions for single-term potentials of the form Eq.~(\ref{eq:v1T}) at zero energy (see the Appendix~\ref{sec:fcp1tDeriv}), we show in the next few subsections that depending on the exponent $\alpha$ and the strength parameter $\pm D_{\alpha}$, all single-term potentials can be grouped into 4 classes, with distinctive cusp functions and strict cusp functions for each group. They are the generalized Coulomb (GC) class with $\alpha<2$, the allowed charge-dipole (alCD) class with $\alpha=2$ and a condition on the strength parameter, the repulsive van der Waals (rVdW) class with $\alpha>2$ and a $+D_{\alpha}$. All other potentials with a single-term singularity, including in particular the attractive van der Waals (aVdW) potentials with $\alpha>2$, are nonphysical. This classification will lead to corresponding classifications of both 2-body and $(N>2)$-body quantum systems according to the short-range behavior of their interaction potentials.

\subsubsection{The generalized Coulomb (GC) class}

Potentials within the generalized Coulomb (GC) class include all single-scale potentials, Eq.~(\ref{eq:v1T}), with $\alpha<2$ including zero and negative $\alpha$'s. The cusp function for the GC class is obtained from the zero-energy solution of Eq.~(\ref{eq:rschuCusp1}), specialized to a single-scale potential [see Eq.~(\ref{eq:rschCusp1s}) of Appendix~\ref{sec:fcp1tDeriv}]. The solution can be viewed from different perspectives that provide different insights. In one perspective, the repulsive and attractive potentials are treated separately. For the ``attractive'' GC (aGC) class, $v^{(\alpha)}=-D_{\alpha}/r^{\alpha}$, we obtain (see Appendix~\ref{sec:fcp1tDeriv})
\begin{equation}
f^{cp(\alpha-)}_{\ell}(r) \overset{\alpha<2}{=} b_{\ell}
	\sqrt{\frac{2}{(2-\alpha)}}r_s^{1/2}J_{\nu_0}(y)\;,
\label{eq:fcpaGC}
\end{equation}
where $b_{\ell}$ is a normalization constant, $J_\nu(y)$ is the Bessel function of the first kind \cite{Olver10},
\begin{equation}
r_s := r/\beta_{\alpha} \;,
\end{equation}
is a scaled, dimensionless, radius,
\begin{multline}
y := \frac{2}{|\alpha-2|}r_s^{(2-\alpha)/2} 
	\overset{\alpha<2}{=} \frac{2}{(2-\alpha)}r_s^{(2-\alpha)/2} \\
	= \frac{2}{(2-\alpha)}\left(\sqrt{\frac{2\mu D_{\alpha}}{\hbar^2}}\right)r^{(2-\alpha)/2}\;,
\end{multline}
and
\begin{equation}
\nu_0 := \frac{2}{|\alpha-2|}(\ell+1/2) 
\overset{\alpha<2}{=} \frac{2}{(2-\alpha)}(\ell+1/2) \;.
\end{equation}

For the ``repulsive'' GC (rGC) class, $v^{(\alpha)}=+D_{\alpha}/r^{\alpha}$, we obtain (see Appendix~\ref{sec:fcp1tDeriv})
\begin{equation}
f^{cp(\alpha+)}_{\ell}(r) \overset{\alpha<2}{=} 
	b_{\ell}\sqrt{\frac{2}{(2-\alpha)}}r_s^{1/2}I_{\nu_0}(y)\;,
\label{eq:fcprGC}
\end{equation}
where $I_\nu(y)$ is the modified Bessel function of the first kind \cite{Olver10}. The $b_{\ell}$, $y$ and $\nu_0$ are the same as those defined for the ``attractive'' class.

Computationally, the cusp functions, given separately in Eqs.~(\ref{eq:fcpaGC}) and (\ref{eq:fcprGC}) for ``attractive'' and ``repulsive'' GC potentials, respectively, are the most convenient. They show that for a single-scale potential such as the GC class, the strength of the interaction, $D_{\alpha}$ comes into play in the cusp function \textit{solely} through the length scale $\beta_{\alpha}$ defined by Eq.~(\ref{eq:betadef}).

Conceptually, and also for the purpose of comparison with existing results for the Coulomb potential \cite{Kato57,Pack66,Tew08,Nagy10}, it is useful to recognize that the results for the aGC class and the rGC class can be combined into a single solution that include, within it, also the free (F) class solution given earlier in Sec.~\ref{sec:Fclass}. 

Define a strength parameter $G_{\alpha}$ that includes the sign of the potential, as in
\begin{equation}
v^{(\alpha)}(r) = \pm \frac{D_{\alpha}}{r^\alpha} =: \frac{G_{\alpha}}{r^\alpha}\;,
\end{equation}
which also implies that $D_{\alpha}=|G_{\alpha}|$. Further pick the normalization constant $b_{\ell}$ to be
\begin{equation}
b_{\ell} = \frac{(2-\alpha)^{\nu_0+1/2}\Gamma(\nu_0+1)}
	{2^{\ell+1}\Gamma(\ell+3/2)} \;,
\label{eq:bell}
\end{equation}
the aGC and rGC solutions can be combined into a single solution
\begin{equation}
f^{cp(\alpha)}_{\ell}(r, s_{L}) = f^{cp(F)}_{\ell}(r, s_{L})I^a_{\nu_0}(z)\;.
\label{eq:fcpGC}
\end{equation}
Here $f^{cp(F)}_{\ell}$ is the free-particle cusp function given by Eq.~(\ref{eq:fcpF}), and we have defined a variable
\begin{equation}
z := (\tfrac{1}{2}y)^2 = \frac{1}{(2-\alpha)^2}\left(\frac{2\mu G_{\alpha}}{\hbar^2}\right) r^{(2-\alpha)} \;,
\end{equation}
and an \textit{analytic function} of $z$, $I^a_{\nu}(z)$, given explicitly by a Taylor expansion
\begin{subequations}
\begin{equation}
I^{a}_{\nu}(z) = \sum_{j=0}^\infty \frac{\Gamma(\nu+1)}{j!\Gamma(\nu+j+1)}z^j \;.
\end{equation}
It is the \textit{analytic portion} of the corresponding modified Bessel function of the first kind $I_{\nu}(y)$ \cite{Olver10}, related by
\begin{equation}
I_{\nu}(y) = \frac{(\tfrac{1}{2}y)^\nu}{\Gamma(\nu+1)}I^a_{\nu}(z) \;,
\end{equation}
and with a property of
\begin{equation}
I^a_{\nu}(z) \overset{z\to 0}{\sim} 1 \;.
\end{equation}
\end{subequations}
The combined solution, $f^{cp(\alpha)}_{\ell}$ of Eq.~(\ref{eq:fcpGC}), is related to the separated aGC and rGC solutions in a simple way, and contains also the F class solution. Specifically,
\begin{subequations}
\begin{align}
f^{cp(\alpha-)}_{\ell}(r) &= f^{cp(\alpha)}_{\ell}(r, s_{L}=\beta_{\alpha})|_{G_{\alpha}=-D_{\alpha}} \;, \\
f^{cp(F)}_{\ell}(r, s_{L}) &= f^{cp(\alpha)}_{\ell}(r, s_{L})|_{G_{\alpha}=0} \;, \\
f^{cp(\alpha+)}_{\ell}(r) &= f^{cp(\alpha)}_{\ell}(r, s_{L}=\beta_{\alpha})|_{G_{\alpha}=+D_{\alpha}} \;.
\end{align}
\end{subequations}
The combined solution $f^{cp(\alpha)}_{\ell}$ has a conceptual value of showing explicitly that it is an analytic function of $G_{\alpha}$, the strength parameter including sign, at $G_{\alpha}=0$ with a corresponding Taylor expansion. The solution depends on $G_{\alpha}$ only through $z$ which has the property of $z\propto G_{\alpha}$. Such an analytic property implies that for the cusp function at any finite $r$, a GC potential can be turned on adiabatically, its solution can evolve analytically from the free-particle solution, and a solution for an attractive potential can evolve analytically from a solution for a repulsive potential. Such characteristics should not be taken for granted. They do not hold for the VdW class of potentials, to be discussed in a later subsection. The introduction of the ``free'' length scale $s_{L}$ emphasizes the physics that $\beta_{\alpha}$, the length scale associated with the strength parameter $D_{\alpha}=|G_{\alpha}|$, disappears in the free-particle limit of $G_{\alpha}\to 0$.

The combined solution is easier for comparison with existing results. Writing out the first few terms of the expansion for $I^a_\nu$ explicitly, we have 
\begin{multline}
f^{cp(\alpha)}_{\ell}(r, s_{L}) = f^{cp(F)}_{\ell}(r, s_{L})\left[1+\frac{1}{\nu_0+1}z\right. \\
	\left.+\frac{1}{2(\nu_0+2)(\nu_0+1)}z^2 + \cdots \right]\;.
\label{eq:fcpGCexp}
\end{multline}
Specializing to the Coulomb potential with $\alpha=1$ and $G_1=Z_1 Z_2 e^2$, the term to first order in $r$ gives the famous Coulomb cusp condition \cite{Kato57,Pack66}. The higher-order terms extends the cusp function to infinite range. Our result also puts the Coulomb potential, both attractive and repulsive cases, within a much more general GC class that include all single-scale $1/r^{\alpha}$ potentials with $\alpha<2$, including the case $\alpha=0$, corresponding to a constant potential, and the case of $\alpha<0$, corresponding to other regular single-scale potentials. Within this context, we note that $z\propto r^{(2-\alpha)}$ with an $\alpha$-dependent exponent. While the first order term in $z$ is of the order $r$ for a Coulomb potential ($\alpha=1$), it would be of the order of $r^3$ for a linear (Stark) potential corresponding to $\alpha=-1$. 

All potentials within the GC class have the same strict cusp function that is independent of $\alpha$, and is given by that of a free particle
\begin{equation}
F^{cp(\alpha)}_{\ell}(r) = \lim_{r\to 0} f^{cp(\alpha)}_{\ell}(r)
	= F^{cp(F)}_{\ell}(r, s_L)  \;,
\label{eq:FcpGC}
\end{equation}
where $F^{cp(F)}_{\ell}(r, s_L)$ is the strict cusp function for the F class given by Eq.~(\ref{eq:FcpF}). In other words, they all have the familiar $r^{\ell+1}$ type of behavior near the origin. 

The cusp functions for the GC class are not only of interest by themselves. They describe universal behaviors shared by all potentials with the behavior
\[
v(r) \overset{r\to 0}{\sim} v^{(\alpha)}(r) = \pm \frac{D_{\alpha}}{r^\alpha} =: \frac{G_{\alpha}}{r^\alpha}\;,
\]
and with $\alpha<2$. We will label this class of potentials by SR-GC class, where SR stands for short-range. It is the most important class of potentials that includes the Coulomb potential $v(r)=G_1/r$, all regular potentials, and all single-scale-singular potentials satisfying the JPN criterion of Eq.~(\ref{eq:JPc}). For all such potentials,
\begin{equation}
f^{cp}_{\ell}(r) \overset{r<r_0}{\sim} f^{cp(\alpha)}_{\ell}(r) \;,
\label{eq:CuspCSRGC}
\end{equation}
where $r_0$ is the radius within which the potential is well represented by a GC potential. It further implies that
\begin{equation}
F^{cp}_{\ell}(r) = \lim_{r\to 0} f^{cp}_{\ell}(r)
	= F^{cp(F)}_{\ell}(r, s_L)  \;,
\label{eq:sCuspCSRGC}
\end{equation}
meaning all potentials of the SR-GC class have the same strict cusp function as the free particle, with $r^{\ell+1}$ type behavior.

\subsubsection{The allowed charge-dipole (alCD) class}

Potentials within the allowed charge-dipole (alCD) class include all single-term potentials, Eq.~(\ref{eq:v1T}), with $\alpha=2$ and a constraint on the strength parameter such that $\gamma_2<1/4$ for an attractive potential. There is no constraint on the strength parameter for a repulsive potential.

The cusp function for the alCD class is easily found from the familiar F class solution, as the potential only transforms the partial wave quantum number $\ell$ into a ``transformed'' $\ell_t$ which is generally no longer an integer. We obtain
\begin{equation}
f^{cp(2)}_{\ell}(r, s_L) = f^{cp(F)}_{\ell_t}(r, s_{L}) = \frac{1}{2^{\ell_t+\tfrac{1}{2}}\Gamma(\ell_t+\tfrac{3}{2})} (r/s_{L})^{\ell_t+1}\;,
\label{eq:fcpalCD}
\end{equation}
where $s_L$ is again a ``free'' length scale,
\begin{equation}
\ell_t = \sqrt{(\ell+\tfrac{1}{2})^2+\gamma_2}-1/2 \;,
\label{eq:lt2}
\end{equation}
for all repulsive CD potentials, and 
\[
\ell_t = \sqrt{(\ell+\tfrac{1}{2})^2-\gamma_2}-1/2 \;.
\]
for attractive CD potentials with $\gamma_2<1/4$. Similar to the F class, the strict cusp function for the alCD class is the same as the cusp function. Specifically,
\begin{equation}
F^{cp(2)}_{\ell}(r, s_L) = \lim_{r\to 0} f^{cp(2)}_{\ell}(r) = F^{cp(F)}_{\ell_t}(r, s_L)\;.
\label{eq:FcpalCD}
\end{equation}
It is the same as the free-particle strict cusp function except for a transformed, strength-dependent $\ell_t$.

The cusp functions for the alCD class are well behaved at $D_{2}=0$. Both repulsive and attractive potentials can be described by a single analytic function by introducing a strength parameter $G_2$ that includes the sign of the potential, as in 
\begin{equation}
v^{(2)}(r) = \pm \frac{D_{2}}{r^2} =: \frac{G_{2}}{r^2}\;,
\end{equation}
and by redefining 
\begin{equation}
\gamma_2 = 2\mu G_{2}/\hbar^2 \;,
\label{eq:gam2def2}
\end{equation}
with $\ell_t$ given by Eq.~(\ref{eq:lt2}) for both repulsive and attractive potentials. In such a formulation that covers both attractive and repulsive alCD potentials, the criterion of $\gamma_2<1/4$ for an attractive potential translates into the condition of $\gamma_2>-1/4$ in the new definition. The cusp function, at any finite $r$, being analytic at $G_2=0$, implies that the potential can be turned on adiabatically, meaning that the solution can evolve analytically from a non-interacting, free-particle solution.

The cusp functions for the alCD class are not only of interest for a pure alCD potential. They also describe universal behaviors shared by all potentials with the behavior
\begin{equation}
v(r) \overset{r\to 0}{\sim} v^{(\alpha)}(r) = \pm \frac{D_{2}}{r^2} =: \frac{G_2}{r^2}\;.
\label{eq:SRalCDdef}
\end{equation}
We will label this class of potentials as the SR-alCD class, where SR again stands for short-range. It is a class of physical potentials that \textit{violates} the JPN criterion of Eq.~(\ref{eq:JPc}). For all such potentials,
\begin{equation}
f^{cp}_{\ell}(r) \overset{r<r_0}{\sim} f^{cp(2)}_{\ell}(r) \;,
\label{eq:CuspCSRalCD}
\end{equation}
where $r_0$ is the radius within which the potential is well represented by an alCD potential. It further implies that
\begin{equation}
F^{cp}_{\ell}(r) = \lim_{r\to 0} f^{cp}_{\ell}(r)
	= F^{cp(F)}_{\ell_t}(r, s_L)  \;,
\label{eq:sCuspCSRalCD}
\end{equation}
meaning all potentials of the SR-alCD class have the same strict cusp function with a $r^{\ell_t+1}$ type behavior with $\ell_t$ being dependent on the strength parameter.

\subsubsection{The repulsive van der Waals (rVdW) class}

Potentials within the repulsive van der Waals (rVdW) class include all single-scale potentials, Eq.~(\ref{eq:v1T}), with $+D_{\alpha}$ and $\alpha>2$. The cusp function for the rVdW class is found in a way similar to the GC class. We obtain  (see the Appendix~\ref{sec:fcp1tDeriv})
\begin{equation}
f^{cp(\alpha+)}_{\ell}(r) \overset{\alpha>2}{=} \frac{2}{\pi}\frac{1}{\sqrt{(\alpha-2)}}r_s^{1/2}K_{\nu_0}(y)\;,
\label{eq:fcprVdW}
\end{equation}
where $K_\nu(y)$ is the modified Bessel function of the second kind \cite{Olver10},
\begin{multline}
y := \frac{2}{|\alpha-2|}r_s^{(2-\alpha)/2} 
	\overset{\alpha>2}{=} \frac{2}{(\alpha-2)}r_s^{-(\alpha-2)/2} \\
	= \frac{2}{(\alpha-2)}\left(\sqrt{\frac{2\mu D_{\alpha}}{\hbar^2}}\right)r^{-(\alpha-2)/2}\;,
\end{multline}
and
\begin{equation}
\nu_0 := \frac{2}{|\alpha-2|}(\ell+1/2) \overset{\alpha>2}{=} \frac{2}{(\alpha-2)}(\ell+1/2) \;.
\end{equation}
The corresponding strict cusp function is obtained from the cusp function through the strict cusp limit of $r\to 0$, as in Eq.~(\ref{eq:stCuspFcusp}). We note that for the rVdW class, with $\alpha>2$, the limit of $r\to 0$ corresponds to the limit of $y\to\infty$. This is different from the GC class, for which, with $\alpha<2$, the limit of $r\to 0$ corresponds to the limit of $y\to 0$. From the large-$y$ asymptotic behavior of the Bessel function $K_\nu(y)$, we have
\begin{subequations}
\begin{equation}
F^{cp(\alpha+)}_{\ell}(r) = \lim_{r\to 0} f^{cp(\alpha+)}_{\ell}(r) 
	= H^{cp(\alpha)}(r) \;,
\end{equation}
where we have defined, for $\alpha>2$,
\begin{equation}
H^{cp(\alpha)}(r) = \frac{1}{\sqrt{\pi}}r_s^{\alpha/4}
	\exp\left(-\frac{2}{\alpha-2}r_s^{-(\alpha-2)/2}\right) \;.
\end{equation}
\label{eq:FcprVdW}
\end{subequations}
Such a strict cusp function is fundamentally different from the $r^{\ell+1}$ behavior for the GC class or the $r^{\ell_t+1}$ behavior for the alCD class, with two unique characteristics. First, the strict cusp function, $F^{cp(\alpha+)}_{\ell}(r)$ for $\alpha>2$, is $\ell$-independent. It is a key characteristic that underlies the partial-wave-insensitive formulations of QDTs \cite{Gao01,Gao08a,Gao20b}, which is an important approach to relate and to treat contributions from a large number of partial waves. Second, unlike the GC class or the alCD class, the cusp function for rVdW class, Eq.~(\ref{eq:fcprVdW}), is singular at $D_{\alpha} = 0$ for any finite $r$. 
It is a reflection that the solution for a rVdW potential is qualitatively different from the free-particle solution ($D_{\alpha}=0$) and solutions for attractive van der Waals (aVdW) potentials, which, as to discussed the next subsection, are nonphysical. Unlike the GC class or the alCD class, a rVdW potential cannot be turned on adiabatically. Its solution does not evolved analytically from the F class of solutions nor the aVdW class of solutions. The rVdW class of potentials violate the JPN criterion of Eq.~(\ref{eq:JPc}). In 2-body scattering by a pure rVdW potential (see, e.g, \cite{Gao99a}), this violation can manifest through a violation of the Levinson's theorem \cite{Gao17a}.

The cusp functions for the rVdW class are not only of interest for a pure rVdW potential. They also describe universal behaviors shared by all potentials with the behavior
\[
v(r) \overset{r\to 0}{\sim} v^{(\alpha)}(r) = + \frac{D_{\alpha}}{r^\alpha} \;,
\]
with $\alpha>2$. We label this class of potentials as the SR-rVdW class. It is a class of potentials that violates the JPN criterion of Eq.~(\ref{eq:JPc}). For all such potentials,
\begin{equation}
f^{cp}_{\ell}(r) \overset{r<r_0}{\sim} f^{cp(\alpha+)}_{\ell}(r) \;,
\label{eq:CuspCSRrVdW}
\end{equation}
where $r_0$ is the radius within which the potential is well represented by a rVdW potential. It further implies that
\begin{equation}
F^{cp}_{\ell}(r) = \lim_{r\to 0} f^{cp}_{\ell}(r)
	= H^{cp(\alpha)}(r)  \;,
\label{eq:sCuspCSRrVdW}
\end{equation}
which means that all potentials of the SR-rVdW class have the same strict cusp function with an $\ell$-independent behavior given by $H^{cp(\alpha)}(r)$. It is the most important class of physical potentials \textit{not} covered by the JPN criterion of Eq.~(\ref{eq:JPc}).

\subsubsection{Nonphysical classes and other classes}
\label{sec:nonphysical}

\begin{table*}
\caption{Classification of physically-allowed 2-body quantum systems according to the short-range behavior of their interaction potential, and a summary of the properties of their corresponding cusp functions. A similar classification applies to physically-allowed ($N>2$)-body system except that they may have multiple classes simultaneously if consisting of particles of multiple kinds. Note that the the strict cusp function for the SR-GC class is independent of $\alpha$, and the strict cusp function for the SR-rVdW class is independent of $\ell$. $r_0$ is the radius within which the potential is well represented by its asymptotic behavior used for classification. It can be infinite if the asymptotic potential is applicable for all $r$.
\label{tb:SRclasses}}
\begin{ruledtabular}
\begin{tabular}{cccc}
Class & Potential & Cusp function property & Strict cusp function  \\
\hline
SR-GC & $v(r) \overset{r\to 0}{\sim} G_{\alpha}/r^\alpha, \alpha<2$ 
	&  $f^{cp}_{\ell}(r) \overset{r<r_0}{\sim} f^{cp(\alpha)}_{\ell}(r, s_L)$ 
	& $F^{cp(F)}_{\ell}(r, s_L)$ \\
SR-alCD & $v(r) \overset{r\to 0}{\sim} G_{2}/r^2, \gamma_2>-1/4$ 
	&  $f^{cp}_{\ell}(r) \overset{r<r_0}{\sim} f^{cp(2)}_{\ell}(r, s_L)$ 
	& $F^{cp(F)}_{\ell_t}(r, s_L)$ \\
SR-rVdW & $v(r) \overset{r\to 0}{\sim} +D_{\alpha}/r^\alpha, \alpha>2$ 
	&  $f^{cp}_{\ell}(r) \overset{r<r_0}{\sim} f^{cp(\alpha+)}_{\ell}(r)$ 
	& $H^{cp(\alpha)}(r)$ \\
SR-alImtS & $v(r) \overset{r\to 0}{\sim} v^{(\{\alpha\})}(r)$
	&  $f^{cp}_{\ell}(r) \overset{r<r_0}{\sim} f^{cp(\{\alpha\})}_{\ell}(r)$ 
	& $F^{cp(\{\alpha\})}_{\ell}(r)$ \\
\end{tabular}
\end{ruledtabular}
\end{table*}

Other than the SR-GC class, the SR-alCD class, and the SR-rVdW class of potentials, all other potentials with a single-term singularity are nonphysical. They include the short-range attractive van der Waals (SR-aVdW) class of potentials that behave as 
\[
v(r) \overset{r\to 0}{\sim} v^{(\alpha)}(r) = - \frac{D_{\alpha}}{r^\alpha} \;,
\]
with $\alpha>2$, and the short-range nonphysical charge-dipole (SR-npCD) class of potentials that behave as Eq.~(\ref{eq:SRalCDdef}) but is too attractive as characterized by $G_2<-1/4$. 
Such potentials being nonphysical are known \cite{LandauQM}, and can be argued from different perspectives. One is such a system would not be stable since it would have an energy spectrum that is not bounded from below \cite{LandauQM}. (See, e.g., Refs.~\cite{LeRoy70,Gao99b,Gao03} for further details on related energy spectra.) The second argument is that the particle(s) would fall to the center in such a system \cite{LandauQM}. The same arguments also apply to potentials with a multi-term singularity if the dominant term is of an aVdW or an npCD class. There are other arguments that can be made to further solidify why such potentials are nonphysical. Because they are lengthy and do not change the conclusion, we will leave them to a better and more relevant context. Here we only emphasize that the same aVdW or npCD type of potentials, which characterize behaviors not allowed near the origin, constitute some of the most important classes of long-range potentials in a QDT \cite{Gao08a}. There is no contradiction. Most physical potentials evolve from one type of physically-allowed behavior near the origin to a different type of behavior at large $r$.

The only other physically-allowed potentials with short-range behaviors not yet classified are those that belong to the allowed intrinsically multi-term-singular (alImtS) class of potentials. They are potentials that behave as Eq.~(\ref{eq:vuo}), with the dominant term being allowed, namely not an aVdW nor a npCD type, but with a behavior that cannot be fully approximated by a single-term solution. Their cusp functions and behaviors, determined by Eqs.~(\ref{eq:rschuCusp1})-(\ref{eq:uCuspC}), will generally require a numerical or a semiclassical solution.

Table~\ref{tb:SRclasses} summaries a classification of physically-allowed 2-body quantum systems according to the short-range behaviors of their interaction potential and the properties of their corresponding cusp functions. From the results of the next section,  it will become clear that a similar classification applies to physically-allowed ($N>2$)-body systems except that they may have multiple classes present simultaneously if consisting of particles of multiple kinds.

\section{2-particle separability and 2-particle cusp conditions in an $N$-body quantum system}
\label{sec:Nbody}

We establish here rigorously the 2-particle separability in an $N$-body quantum system. Specifically we show that a 2-body subsystem within an $(N>2)$-body system becomes separable from the rest of the system in a version of the coalescence limit $r\to 0$ that is generally different from the strict cusp limit of the previous section. The 2-particle cusp conditions in an $N$-body quantum system follow naturally from separability, but require an understanding of the energy and the partial wave dependences of the 2-body solution at the level of the previous section. Both results are part of a reconstructed foundation of an $N$-body quantum theory. Through their derivations we hope to explain the subtleties that, until now, may have prevented their understanding.

\subsection{2-particle separability in an $N$-body system}

Consider an $N$-body quantum systems with $N>2$ interacting via pairwise potentials. It is described by a Hamiltonian
\begin{equation}
H^{(N)} = \sum_{i=1}^N -\frac{\hbar^2}{2m_i}\nabla_i^2 + \sum_{i<j=1}^N \hat{v}_{ij}(r_{ij})\;,
\label{eq:nbH}
\end{equation}
and a corresponding Schr\"odinger equation at an energy $E$
\begin{equation}
H^{(N)}\Psi^{(N)}_{E\gamma} = E\Psi^{(N)}_{E\gamma} \;.
\end{equation}
Here $\gamma$ represents all quantum numbers required to uniquely specify an $N$-body state. $\hat{v}_{ij}(r_{ij})$ is the interaction potential between particles $i$ and $j$. It is generally an operator if more-than-one internal states, including spin states, are involved in the energy range of interest. For simplicity, but without loss of generality for our purposes here, we assume $\hat{v}_{ij}(r_{ij}) = v_{ij}(r_{ij})$ to be a scalar central potential. The cases of spin-dependent, multichannel, and/or anisotropic potentials \cite{Gao20b} will be addressed in future studies.

For any chosen pair $(i,j)$ out of the $N>2$ particles, which may, or may not, be identical, the Hamiltonian can always be regrouped, using the center-of-mass frame of the pair $(i,j)$, as
\begin{equation}
H^{(N)} = h_{ij}+H^{(ij)} \;.
\end{equation}
Here
\begin{equation}
h_{ij} := -\frac{\hbar^2}{2\mu_{ij}}\nabla_r^2+v_{ij}(r) \;,
\end{equation}
describes the relative motion of the pair $(i,j)$ interacting via a two-body central potential $v_{ij}(r)$, in the center-of-mass frame of the pair $(i,j)$ defined by 
\begin{subequations}
\begin{align}
\bm{r} &:= \bm{r}_{ij} := \bm{r}_j-\bm{r}_i \;,\\
\bm{c} &:= (m_i\bm{r}_i+m_j\bm{r}_j)/(m_i+m_j) \;,
\end{align}
\end{subequations}
and with $\mu_{ij} = m_i m_j/(m_i+m_j)$ being the 2-particle reduced mass. $H^{(ij)}$ is the Hamiltonian for all other degrees of freedom that include the other $N-2$ particles, the center-of-mass motion of the pair $(i,j)$, and the interaction between pair $(i,j)$ and other particles. Specifically,
\begin{equation}
H^{(ij)} := H^{(N-2)}-\frac{\hbar^2}{2 m_{ij}}\nabla_c^2+V^{(ij)} \;,
\end{equation}
where $H^{(N-2)}$ is the Hamiltonian describing the $(N-2)$-body subsystem, $-\tfrac{\hbar^2}{2 m_{ij}}\nabla_c^2$ with $m_{ij}=m_i+m_j$ is the kinetic energy of the center-of-mass of the pair $(i,j)$, and
\begin{equation}
V^{(ij)} := \sum_{k\neq(i,j)}^N \left[v_{ik}(r_{ik})
	+ v_{jk}(r_{jk}) \right] \;,
\end{equation}
is the interaction between the pair $(i,j)$ and the $(N-2)$-body subsystem.

The first key to deriving cusp conditions for an $N$-body system is the recognition that the two-body relative motion of the pair $(i,j)$ become separable from the other degrees of freedom in the limit of $r\to 0$. Specifically, we have, for basically all physical potentials (see Appendix~\ref{sec:SepNotes})
\begin{equation}
V^{(ij)} \overset{r\to 0}{\sim} V^{(ij)}_{sp} 
	:= \sum_{k\neq(i,j)}^N \left[v_{ik}(R_k)
	+ v_{jk}(R_k) \right] \;,
\label{eq:NbVsep}
\end{equation}
becoming independent of $\bm{r}$, where we have defined $R_k := |\bm{r}_k-\bm{c}|$ being the distance from particle $k$ to the center-of-mass of the pair $(i,j)$. Correspondingly, we have
\begin{equation}
H^{(ij)} \overset{r\to 0}{\sim} H^{(ij)}_{sp} := H^{(N-2)}-\frac{\hbar^2}{2 m_{ij}}\nabla_c^2+V^{(ij)}_{sp} \;,
\end{equation}
becoming independent of $\bm{r}$, and
\begin{equation}
H^{(N)} \overset{r\to 0}{\sim} H^{(N)}_{sp} := h_{ij}+H^{(ij)}_{sp} \;,
\label{eq:NbHsep}
\end{equation}
becomes two parts that are independent of each other. We are using the subscript of $sp$ to represent the results in the separable limit. 

The separability of the pair $(i,j)$ relative motion, described by $h_{ij}$, from the other degrees of freedom implies that all 2-body properties that we have discussed in Sec~\ref{sec:2body}, including the physical boundary condition of Eq.~(\ref{eq:PBC3D}) leading to the 2-particle coalescence being a naturally pinned point, and the resulting concept of rigidity and all its consequences, apply to the pair $(i,j)$ relative motion in an $N$-body system for a sufficiently small $r$. Mathematically, the separability further implies that in the limit of $r\to 0$, the $N$-body wave function can always be written as a linear superposition of product states, specifically,
\begin{multline}
\Psi^{(N)}_{E\gamma} \overset{r\to 0}{\sim} 
	\sum_{\lambda\ell m_\ell s m_s\gamma'} a_{\lambda\ell m_\ell s m_s\gamma'}
	\psi_{\lambda\ell m_\ell s m_s}
	\Psi^{sp}_{(E-\lambda)\gamma'} \;,\\
	\overset{r\to 0} {\sim} 
	\sum_{\lambda\ell m_\ell s m_s\gamma'} a_{\lambda\ell m_\ell s m_s\gamma'}
	\frac{u_{\lambda\ell}(r)}{r}Y_{\ell m}(\theta,\phi)|sm_s\rangle
	\Psi^{sp}_{(E-\lambda)\gamma'} \;.
\label{eq:NbWfnSep}
\end{multline}
Here $\psi_{\lambda\ell m_\ell s m_s}$ is the wave function for the pair $(i,j)$ relative motion at a pair energy $\lambda$,
\begin{equation}
h_{ij}\psi_{\lambda\ell m_\ell s m_s} 
	=  \lambda\psi_{\lambda\ell m_\ell s m_s} \;,
\end{equation}
with $u_{\lambda\ell}(r)$ being its radial component, $\bm{s}=\bm{s}_i+\bm{s}_j$ being the total spin of the pair, and $m_s$ being its projection on a $z$-axis. $\Psi^{sp}_{(E-\lambda)\gamma'}$ is an eigenfunction of $H^{(ij)}_{sp}$ at energy $E-\lambda$, for the other $N-2$ particles plus the  center-of-mass motion of the pair $(i,j)$,
\begin{equation}
H^{(ij)}_{sp}\Psi^{sp}_{(E-\lambda)\gamma'}
	= (E-\lambda)\Psi^{sp}_{(E-\lambda)\gamma'} \;.
\end{equation}
The sum over direct products, with $a_{\lambda\ell\dots}$ being the constant coefficients of superposition, is constrained by symmetries and conservation laws, with details determined by the degree to which the 2-particle states are mixed by its interaction with other $(N-2)$ particles \textit{outside} of the region of separability.  In particular, the sum over the pair $(i,j)$ relative energy $\lambda$, which generally includes both a sum over a discrete spectrum and an integration over a continuum spectrum for a pair $(i,j)$ that can bind, describes a mixing of pair wave functions of different energies. At a simplest qualitative level, this mixing can be characterized by a mean $\bar{\lambda}$, and an uncertainty $\Delta\lambda$ which gives a measure of the degree of mixing. We call Eq.~(\ref{eq:NbWfnSep}), a rigorous result on an $N$-body wave function, the separability condition, more precisely a 2-particle separability condition of an $(N>2)$-body wave function. It is a rigorous mathematical statement of a physical picture that we, and perhaps others, have envisioned many years ago \cite{Gao03}.

Both the separability and the resulting wave function separability condition are straightforward. The mathematics behind the separability, Eq,~(\ref{eq:NbHsep}), at least that specialized to a Coulomb potential, is contained, if somewhat hidden and not stated explicitly, in the derivations of the Coulomb cusp conditions, in particular those by Pack and Brown \cite{Kimball73} and by Kimball \cite{Kimball73}. The wave function separability condition of Eq.~(\ref{eq:NbWfnSep}) is an automatic consequence. The idea of separability has motivated multiple significant works in many-body physics \cite{Kimball75,ZL09,Werner12a,Hofmann13}, but in our opinion, never been fully established nor fully understood. The reason is likely twofold. One is that without an understanding of the energy dependence of the pair wave function presented in Sec.~\ref{sec:2body}, not much conclusion can be drawn from the separability condition of Eq.~(\ref{eq:NbWfnSep}) due to the energy mixing. The alternative approaches that avoided dealing directly with energy \cite{Pack66,Kimball73,Kimball75} made use of the Coulomb potential being singular at the coalescence, leading to an impression and/or a belief that a dominant singular potential was the key to separability. In reality, our derivation shows that the 2-particle relative motion in a $N$-body system is always separable, for sufficiently small $r$, from the other degrees of freedoms for all physical potentials that is continuous in $(0,\infty)$ and once or twice differentiable (see Appendix~\ref{sec:SepNotes}). 

The other subtlety, necessary to fully understand and extract the consequences of separability, is a deeper understanding of the separable limit, represented by $r\to 0$ in Eq.~(\ref{eq:NbHsep}), specifically its potential difference from the strict cusp limit, also represented by $r\to 0$ in the 2-body context. In the 2-body theory of Sec.~\ref{sec:2body}, the limit of $r\to 0$, under which we have established the strict cusp conditions, means more rigorously $r/\beta_{\alpha}\to 0$ where $\beta_{\alpha}$ is the shortest of length scales in the potential, or $r/s_L\to 0$ if the potential itself, being F or alCD type, has no length scale. This mixture of cases is part of the reason, in addition to simplicity, that we have chosen the $r\to 0$ notation for the strict cusp limit. It would be easy to think both $r\to 0$'s mean the same. If this were true, only strict cusp conditions would leave their traces in an $N$-body system.

In a 2-body system, the only intrinsic length scales are those associated with the potential, namely the $\beta_{\alpha}$'s. In transition from 2- to 3-body and beyond, a new length scale, $r_{\rho}$, emerges. It is the average value of $R_k = |\bm{r}_k-\bm{c}|$ in Eq.~(\ref{eq:NbVsep}), namely the mean separation of the center-of-mass of pair $(i,j)$ from other particles. The precise value of $r_{\rho}$ does not matter, what is important are its qualitative features. It is an independent length scale well defined starting from a 3-body system. It has an order of magnitude of $r_{\rho}= (4\pi\rho_{\#}/3)^{-1/3}$ for a many-body system in terms of local number density $\rho_{\#}$  (the number of particles per unit volume). It can be as small as $\beta_{\alpha}$ for a tightly-bound few-body system or a dense many-body system, and can be infinitely large, e.g., in a weakly-bound few-body system, or in the low-density limit of a many-body system.

With the emergence of $r_{\rho}$, the coalescence limit of $r\to 0$ has now two distinctive possibilities of $r/\beta_{\alpha}\to 0$ and $r/r_{\rho}\to 0$. Though their mathematical meanings are the same for a tightly-bound few-body states or a high-density many-body states for which $r_{\rho}\sim\beta_{\alpha}$, they can be very different for loosely-bound few-body states and low-density many-body states, for which we can have $r_{\rho}\gg\beta_{\alpha}$ with no upper limit. To determine which of the two limits represents the separable limit, one needs a closer look at the correction terms beyond the separable limit of $H^{(N)}$, Eq.~(\ref{eq:NbHsep}), at terms of higher order in $r$.

For all $N$-body systems with electromagnetic-type interactions, which include all atomic and molecular interactions, we show in Appendix~\ref{sec:SepNotes} that the separable limit is more precisely $r/r_{\rho}\to 0$, or characterized as an asymptotic region as the region of $r\ll r_{\rho}$. Since systems of electromagnetic-type are our primary focus, we rewrite, explicitly for such systems, the separable limit and the wave function separability condition more precisely as
\begin{equation}
H^{(N)} \overset{r\ll r_{\rho}}{\sim} h_{ij}+H^{(ij)}_{sp} \;,
\label{eq:NbHsepEM}
\end{equation}
and
\begin{multline}
\Psi^{(N)}_{E\gamma} \overset{r\ll r_{\rho}}{\sim} 
	\sum_{\lambda\ell m_\ell s m_s\gamma'} a_{\lambda\ell m_\ell s m_s\gamma'}
	\psi_{\lambda\ell m_\ell s m_s}
	\Psi^{sp}_{(E-\lambda)\gamma'} \;,\\
	\overset{r\ll r_{\rho}}{\sim} 
	\sum_{\lambda\ell m_\ell s m_s\gamma'} a_{\lambda\ell m_\ell s m_s\gamma'}
	\frac{u_{\lambda\ell}(r)}{r}Y_{\ell m}(\theta,\phi)|sm_s\rangle
	\Psi^{sp}_{(E-\lambda)\gamma'} \;.
\label{eq:NbWfnSepEM}
\end{multline}
Thus for electromagnetic-type systems at sufficiently low densities, we can have the separable condition of $r\ll r_{\rho}$ well satisfied while simultaneously having $r\gg\beta_{\alpha}$, namely not only outside of the region of strict cusp condition but also outside of the range of the 2-body potential entirely. This understanding of separability, in terms of the emergent length scale $r_{\rho}$, is one of the keys to future formulations of $N$-body theories, including the cusp conditions of the next subsection. And with the introduction of $r_{\rho}$, the degree of mixing of 2-particle energy states, as measured by $\Delta\lambda$, can be estimated to be of the order of $v_{ik}(r_{\rho})$, or $v_{jk}(r_{\rho})$ if different and greater in magnitude.

\subsection{2-particle cusp conditions in an $N$-body system}
 
We focus, hereafter, on systems of electromagnetic type. The separability condition of Eq.~(\ref{eq:NbWfnSepEM}) makes it clear how aspects of 2-body correlation, as reflected in $u_{\lambda\ell}$, survive in an $(N>2)$-body system. Specifically, it is the energy and the partial wave dependences of $u_{\lambda\ell}$ that determine how they survive and are reflected in an $N$-body solution. From Eq.~(\ref{eq:uelcuspC}), or Eq.~(\ref{eq:ue}) together with the property of $x^{(i)}_\ell$ given by Eq.~(\ref{eq:xlr0}), we have from Eq.~(\ref{eq:NbWfnSepEM})
\begin{equation}
\Psi^{(N)}_{E\gamma} \overset{r\ll r_{\rho}}{\sim} \sum_{\ell} \Phi^{(N)}_{E\gamma\ell}
	\frac{C^{cp}_{E\gamma\ell}f^{cp}_{\ell}(r)}{r} \;.
\label{eq:NbCuspC}
\end{equation}
Here $f^{cp}_{\ell}(r)$ is the 2-particle cusp function defined earlier for an arbitrary pair potential. The $\Phi^{(N)}_{E\gamma\ell}$, defined by
\begin{equation}
C^{cp}_{E\gamma\ell}\Phi^{(N)}_{E\gamma\ell}
	= \sum_{\lambda m_\ell s m_s\gamma'} a_{\lambda\ell m_\ell s m_s\gamma'}
	c^{cp}_{\lambda\ell}\Psi^{sp}_{(E-\lambda)\gamma'}
	Y_{\ell m}(\theta,\phi)|sm_s\rangle \;,
\end{equation}
is a wave function describing all degrees of freedom of the $N$-body system other than $r$, and we can choose it to be normalized such that
\begin{equation}
\langle \Phi^{(N)}_{E\gamma'\ell'}|\Phi^{(N)}_{E\gamma\ell}\rangle 
	= \delta_{\gamma'\gamma}\delta_{\ell'\ell} \;.
\end{equation}
$C^{cp}_{E\gamma\ell}$ is an energy-dependent constant that we call the \textit{2-particle (partial) cusp amplitude for an $N$-body quantum system}. Just like the 2-particle (partial) cusp amplitude, $c^{cp}_{\lambda\ell}$, it is a global property that depends on conditions beyond the cusp region, which in $(N>2)$-body cases include interactions of the pair $(i,j)$ with other particles. Its differences from the two-body cusp amplitude contain information on the effects of other particles in an $N$-body system on the short-range pair correlation. 

Equation~(\ref{eq:NbCuspC}) is what we call the general \textit{2-particle cusp condition for an $N$-body quantum system}. It is a general asymptotic behavior in the separable region of $r\ll r_{\rho}$, in the following sense. For any $N$-body system in any state, it is satisfied for sufficiently small $r$. Importantly and similar to the 2-particle cusp condition in a 2-body system, Eq.~(\ref{eq:NbCuspC}) can be applicable over a wide range of $r$, depending on the density and the energy (temperature). For any finite $r$, including those outside of the range of interaction, namely $r>\beta_n$ (see Appendix~\ref{sec:SepNotes}), Eq.~(\ref{eq:NbCuspC}) is satisfied at sufficiently low energies (temperatures) corresponding to sufficiently small $\bar{\lambda}$'s and sufficiently low densities corresponding to sufficiently small $\Delta\lambda$'s. Since the $f^{cp}_{\ell}(r)$ in Eq.~(\ref{eq:NbCuspC}) is same cusp function defined for a 2-body system, an $(N>2)$-body system follows the same short-range universal behaviors and can be classified the same as a 2-body system, as in Table~\ref{tb:SRclasses}.

Contained in Eq.~(\ref{eq:NbCuspC}) is the physics that the pair wave functions in different partial waves are generally mixed due to interactions with other particles. This mixing is generally important for tightly-bound few-body states and for many-body systems of high densities. It is worth noting, however, that there are important special cases in which the interaction and the state of interest is such that $\ell$ is conserved or approximately conserved. This is especially true for a pair $(i,j)$ that can attract and bind. For a bound pair $(i,j)$, no matter how weakly, its relative angular momentum $\ell$ is approximately conserved if its size is much smaller than $r_{\rho}$, its mean separation from other particles. This can happen in a few-body system if the other particles are more loosely bound, and in a many-body system if the density of other particles is sufficiently small. We write out the 2-particle cusp condition for such cases more explicitly as
\begin{equation}
\Psi^{(N)}_{E\gamma} \approx \Psi^{(N)}_{E\gamma\ell} \overset{r\ll r_{\rho}}{\sim} \Phi^{(N)}_{E\gamma\ell}\frac{C^{cp}_{E\gamma\ell}f^{cp}_{\ell}(r)}{r} \;.
\label{eq:NbCuspCl}
\end{equation}
The ability to describe such 2-particle ``molecular'' states naturally and self-consistently within an $N$-body theory is a first important step towards a better understanding of $N$-body systems made of particles that can bind. 

Also contained in Eq.~(\ref{eq:NbCuspC}) are more strict 2-particle cusp condition in an $N$-body system, applicable over a more restricted range of $r$ and obtained by considering further limits towards the eventual strict cusp limit of $r\to 0$. For the SR-GC class and SR-alCD class of systems, $f^{cp}_{\ell}(r)$ follows universal behavior described by $f^{cp(\alpha)}_{\ell}(r)$ within a $r_0$, with $\alpha<2$ for the SR-GC class and $\alpha=2$ for SR-alCD class (see Table~\ref{tb:SRclasses}). These $f^{cp(\alpha)}_{\ell}(r)$'s depend on the partial wave $\ell$ with a predominant dependence of the type $r^\ell$ or $r^{\ell_t}$ as in Eqs.~(\ref{eq:fcpGC}) and (\ref{eq:fcpalCD}). For any $\ell$-mixture states not included in Eq.~(\ref{eq:NbCuspCl}), the sum over $\ell$ becomes, at sufficiently small $r$, dominated by the term with smallest $\ell$. We obtain, 
\begin{subequations}
\begin{align}
\Psi^{(N)}_{E\gamma} {\sim}& \Phi^{(N)}_{E\gamma\ell_{cp}}
	\frac{C^{cp}_{E\gamma\ell_{cp}}f^{cp(\alpha)}_{\ell_{cp}}(r, s_L)}{r} \;, 
	\label{eq:NbsCuspCGCalCD1}\\
	\overset{r\to 0}{\sim}& \Phi^{(N)}_{E\gamma\ell_{cp}}
	\frac{C^{cp}_{E\gamma\ell_{cp}}F^{cp(F)}_{\ell_{cp}}(r, s_L)}{r} \;.
	\label{eq:NbsCuspCGCalCD2}
\end{align}
\label{eq:NbsCuspCGCalCD}
\end{subequations}
Here $\ell_{cp}$ is the smallest $\ell$, or $\ell_t$ for the SR-alCD class, allowed by symmetry in the summation of Eq.~(\ref{eq:NbCuspC}), and $F^{cp(F)}_{\ell_{cp}}$ is the strict cusp function given by Eq.~(\ref{eq:FcpF}). Contained in Eq.~(\ref{eq:NbsCuspCGCalCD1}) is the special case ($\alpha=1$) of the Coulombic interaction that can be compared to the traditional results \cite{Kato57,Pack66}. 
For electrons in a usual atomic or molecular state, $r_{\rho}$ and $\beta_{\alpha}$ are comparable, both of the order of 1 bohr (atomic unit). The separable constraint, $r\ll r_{\rho}$, limits the  range of applicability of the cusp conditions to $r\ll\beta_{\alpha}$, which in turn limits the number of useful terms in the expansion of $f^{cp(\alpha)}_{\ell_{cp}}$, Eq.~(\ref{eq:fcpGCexp}). More precisely in this case, the $N$-body corrections and energy corrections will come in at the order of $r^2$ and become entangled with the $r/\beta_{\alpha}$ expansion of the cusp function, as demonstrated in other excellent efforts on higher orders corrections \cite{Nagy01,Tew08,Nagy10}. This does not mean that the cusp function to infinity range, as given by Eq.~(\ref{eq:fcpGC}), is not useful in the Coulomb case ($\alpha=1$), it simply means that it needs a proper context of ``low-density'' states.

For the SR-rVdW class of systems, using Eq.~(\ref{eq:FcprVdW}), we obtain in the strict cusp limit
\begin{align}
\Psi^{(N)}_{E\gamma} \overset{r/\beta_{\alpha}\to 0}{\sim}& 
	\left(\sum_{\ell}\Phi^{(N)}_{E\gamma\ell}C^{cp}_{E\gamma\ell}\right)
	\frac{H^{cp(\{\alpha\})}(r)}{r} \;,\\
	\overset{r/\beta_{\alpha}\to 0}{\sim}& \Phi^{(N)}_{E\gamma}
	\frac{C^{cp}_{E\gamma}H^{cp(\{\alpha\})}(r)}{r} \;.
	\label{eq:NbsCuspCrVdW}
\end{align}
Here $\Phi^{(N)}_{E\gamma}$ and $C^{cp}_{E\gamma}$ can be defined such that
\begin{equation}
\Phi^{(N)}_{E\gamma} = \left(\sum_{\ell}\Phi^{(N)}_{E\gamma\ell}
	C^{cp}_{E\gamma\ell}\right)\left(C^{cp}_{E\gamma}\right)^{-1} \;,
\end{equation}
and
\begin{equation}
\langle \Phi^{(N)}_{E\gamma'}|\Phi^{(N)}_{E\gamma}\rangle = \delta_{\gamma'\gamma} \;.
\end{equation}
$C^{cp}_{E\gamma}$ is an energy-dependent constant that can be called the \textit{2-particle (total) cusp amplitude for an $N$-body quantum system} to distinguish it from $C^{cp}_{E\gamma\ell}$ when necessary.  

Equations~(\ref{eq:NbsCuspCGCalCD}) and (\ref{eq:NbsCuspCrVdW}) are what we call the \textit{strict 2-particle cusp condition for an $N$-body quantum system}. They are simpler and represent the shorter-range universal behaviors embedded in the more general cusp condition of Eq.~(\ref{eq:NbCuspC}). 

\section{Discussions}
\label{sec:disc}

\subsection{Beyond pairwise potentials}

While we have assumed pairwise potentials for $(N>2)$-body systems for definiteness and for initial applications, it is important to point out that our results only require that the total $N$-body potential $V^{(N)}$ satisfies
\begin{equation}
V^{(N)}(\bm{r}_1,\cdots,\bm{r}_N) \overset{r\ll r_{\rho}}{\sim} v_{ij}(r) + V^{(ij)}_{sp} \;,
\label{eq:VsepCond}
\end{equation}
with $V^{(ij)}_{sp}$ being independent of $r=r_{ij}$. Here $v_{ij}(r)$ can furthermore be different from the pair potential in the absence of other particles. In other words, nonadditive potentials would lead to the same conclusion provided they satisfy the condition of Eq~(\ref{eq:VsepCond}). It is a much weaker condition than pairwise potentials.  Mathematically, it only requires that $V^{(N)}$ is well behaved everywhere with only possible exceptions at the points of coalescence, where, if singular, the nature of the singularity depends only on the interparticle separations of the particles approaching coalescence. It is our expectation that such a condition is satisfied for all physical potentials, including nonadditive potentials for systems of composite particles such as atoms (see, e.g., \cite{Stone13}). There is no change to the theory except for a proper reidentification of $v_{ij}(r)$ that defines the cusp functions and cusp conditions.

\subsection{Multiparticle rigidity and multiparticle cusp conditions}

The concept of 2-particle rigidity and 2-particle cusp conditions in 2-body and $(N>2)$-body systems can be extended to $N_{cp}$-particle rigidity and $N_{cp}$-particle cusp conditions in $N_{cp}$-body and $(N>N_{cp})$-body quantum systems with $N_{cp}>2$. The extension can be carried out using a hyperspherical coordinate system (see, e.g., Refs.~\cite{Avery18,Greene17}). Let $R_{cp}$ be the hyperradius for an $N_{cp}$-particle system, one can again show, from the Schr\"odinger equation for the hyperradial degree of freedom \cite{Greene17}, that the $N_{cp}$-particle coalescence point, characterized by $R_{cp}\to 0$, is a pinned point with its related concept of rigidity and $N_{cp}$-particle cusp conditions. Similarly, such cusp conditions survive in a $(N>N_{cp})$-body system because the $N_{cp}$-particle subsystem becomes separable from the the rest of the system in a version of the coalescence limit of $R_{cp}\to 0$.

It is useful to note, however, that a multiparticle cusp condition is qualitatively different, not contained in the 2-particle cusp condition only if the interaction potential $v_{ij}(r)$ is singular at $r=0$. The infinite rigidity at the 2-particle coalescence means that the behavior around it remains unchanged by the presence of other particles if their interactions with those particles remain finite. Even when nontrivial multiparticle cusp conditions do exist, such as for singular potentials, their effects can be expected to be small except at extremely high densities where ultimately all atomic interactions becomes Coulombic \cite{Buckingham58,Nikitin84}. Thus beyond Coulombic systems \cite{Fournais05,Hoffmann92}, the generalization to multiparticle rigidity and multiparticle cusp conditions, while conceptually interesting, is not urgent and is not expected to have much qualitative impact in practice.

\section{Conclusions and outlook}
\label{sec:conc}

In conclusion, we have introduced a concept of rigidity to measure the sensitivity of a wave function to changes in energy, potential, and/or external perturbation. Through this concept, we have established cusp conditions and related functions as fundamental properties of an arbitrary $N$-body quantum system with $N\ge 2$.

In a 2-body system, we have defined, and made distinction of, both the cusp function and the strict cusp function, of which the cusp function is well defined at all interparticle separations and can be found either analytically, when possible, or numerically for any physical potential.
Importantly, we have established the connection between the cusp function and the general analytic properties of a 2-body regular solution versus energy. These general analytic properties, together with the concept of rigidity and its related properties, further strengthen the mathematical and conceptual foundation of 2-body QDTs \cite{Seaton83,Greene79,Greene82,Greene85,Gao08a,Gao20b}, and provide motivations and tools for their further development.

We have derived analytic cusp functions and strict cusp functions for all single-term physical potentials and used them to provide a general classification of quantum systems based on the short-range behavior of their interaction potential. It shows that there are classes of quantum systems, in particular the SR-rVdW class, that have short-range behaviors completely different from the more standard and more familiar $r^{\ell+1}$ type of behavior that characterizes the SR-GC class.

We have shown that the relative motion of any 2-particle subsystem of an ($N>2$)-body system is separable from other degrees of freedom under a separable limit of $r/r_{\rho}\to 0$ which is a version of the coalescence limit that can be substantially different from the strict cusp limit of $r/\beta_{\alpha}\to 0$. From this separability, we have derived the general 2-particle cusp conditions for an arbitrary ($N>2$)-body system and their classifications according to the short-range potential. Such a theory for cusp conditions and functions can be called a QDT for short-range correlation, to reflect its mathematical and conceptual similarity to traditional QDTs focusing on the long-range correlation \cite{Seaton83,Greene79,Greene82,Greene85,Gao08a,Gao20b}.

Amid the complexity of an $N$-body quantum system, the cusp conditions and the rigidity of a wave function around the coalescence points provide important anchor points upon which a quantum theory of strongly interacting particles can be built. We emphasize that similar to the physical boundary condition, Eq.~(\ref{eq:PBC3D}), for a 2-body system, the cusp conditions for an $N$-body system are not ``contained'' in the Schr\"odinger equation and need to be enforced to ensure a correct solution. Without enforcing such conditions, one can hope, but cannot be sure, that a solution is a reasonable approximation to the real physics. This explains the success of quantum Monte Carlo (QMC) methods based on Jastrow and Slater-Jastrow functions \cite{Jastrow55,Foulkes01,Austin12,Gruneis17}. Such calculations are often accurate because their trial wave functions can be chosen, and often have been chosen as inherent in the Jastrow formulation \cite{Jastrow55}, to satisfy the correct cusp conditions. A key forward is to build theories in which the cusp conditions are enforced rigorously and efficiently, not only for a particular state but for all states including highly excited states and scattering and reactive states.

With little or no modification, the results presented here can be used with existing theories of $N$-body quantum systems, such as Hartree-Fock-type mean-field theory or quantum Monte Carlo methods \cite{Jastrow55,Foulkes01,Austin12,Gruneis17} to improve the description of short-range correlation. For systems with purely repulsive potentials \cite{Gao17a}, such improvement is likely already sufficient to provide an accurate description of the system, similar to the improvements of R12 \cite{Klopper87,Kutzelnigg91,Gruneis17} and F12 \cite{Ten-no04,Ten-no07,Knizia09,Gruneis17} theories to the Hartree-Fock theory of electronic structure. 

For systems of particles that can attract and bind, such as systems of atoms and molecules (in which the nuclei are also treated quantum-mechanically unlike in electronic structure), the results and the concepts here will play a more fundamental role in enabling new classes of $N$-body theories. While hints and hopes for such theories can be found in some of our old works \cite{Gao03,Gao04a,Gao05b,KG06}, the mathematical and conceptual foundation were not sufficiently strong nor sufficiently general for their further development at the time. The results presented here are expected to be sufficient to build better theories for at least some phases and/or states of such systems, in particular the states of a single homogeneous phase and states in which only 2-particle binding needs to be considered. We hope to soon illustrate such theories with applications to ultracold atomic or molecular gases, motivated both by their relative simplicity and more importantly by the prospects of comparison with precise experiments \cite{BDZ08,Bohn17}.

For a more complete description of vastly many other states of $N$-body quantum systems made of particles that can bind, including states of mixed phases consisting of clusters or molecules of different sizes, nanoparticles, droplets, and surfaces, we will need at least one more piece of foundation. It is a foundation built around the concept of \textit{arrangement}, known in collisions and reactions \cite{Taylor06} and to be generalized in an upcoming work to an $N$-body quantum system \cite{Gao22b}, to further address the unique complexities and difficulties associated with binding. The concept of arrangement is another emergent concept in going from 2- to 3-body, similar to the the length scale $r_{\rho}$ of this work, but unique for systems of particles that can attract and bind. Also important for such systems will be multiparticle separability, namely an $N_{cp}$-particle subsystem becomes separable from the the rest of the system in a version of the coalescence limit of $R_{cp}\to 0$ (see Sec.~\ref{sec:disc}). These topics will be addressed in an upcoming paper \cite{Gao22b}.

With such a reconstructed foundation, and a better treatment of intermediate and long-range correlations in $N$-body systems using QDT methods that further build upon our old works \cite{Gao03,Gao04a,Gao05b,KG06}, we expect to be in a much closer position to build rigorously what we call QDTs for $N$-body quantum systems, not only for the ground-state structure, but also for excited states and excited structures, and for interactions and reactions. They are expected to reflect more fully our view that the mathematical essence of quantum mechanics is an efficient representation of functions. For a function that we know both its short-range and long-range behaviors sufficiently well to squeeze the unknown into a finite region in space in which the potential is finite and continuous, the unknown should be representable by a polynomial or other equivalent finite basis. This is true for a function of a single variable, and can be expected to remain true for multiparticle wave functions for which the points of coalescence serve as effective origins. This point of view can be partly traced back to the $R$-matrix theory of 2-body interactions pioneer by Wigner and coworkers \cite{Wigner47,TW52,Aymar96,Burke11}, and we intend to push it to an extreme in future QDT formulations, by squeezing on both ends when needed and through generalizations to $N$-body systems.

\begin{acknowledgments}
I thank Dr. Ningyi Du for helpful discussions. This work was supported by NSF under Grant No. PHY-1912489.
\end{acknowledgments}

\appendix

\section{Derivation of the fundamental equations of rigidity}
\label{sec:rgFRderiv}

The fundamental equation of rigidity for 1-D, Eq.~(\ref{eq:rgFR1D}), is obtained from Eq.~(\ref{eq:rgR1D}) and the definition of rigidity, Eq.~(\ref{eq:rgdef1D}). Equation~(\ref{eq:rgR1D}) is derived from the 1-D Schr\"odinger equations at two different energies $\epsilon$ and $\epsilon'$, 
\begin{align}
\left[-\frac{\hbar^2}{2m}\frac{d^2}{dx^2} 
	+ v(x) \right]
	\psi_{\epsilon}(x) &= \epsilon \psi_{\epsilon}(x) \;,\\
\left[-\frac{\hbar^2}{2m}\frac{d^2}{dx^2} 
	+ v(x) \right]
	\psi_{\epsilon'}(x) &= \epsilon' \psi_{\epsilon'}(x) \;,
\end{align}
from which it is straightforward to show
\[
(\epsilon'-\epsilon)\int_0^x \psi_{\epsilon'}(x')\psi_{\epsilon}(x')dx' 
	= \frac{\hbar^2}{2m}\left.\mathscr{W}_{x'}\left(\psi_{\epsilon'}(x'),\psi_{\epsilon}(x')\right)\right|_0^x \;,
\]
or
\begin{equation}
\int_0^x \psi_{\epsilon'}(x')\psi_{\epsilon}(x')dx' 
	= \frac{\hbar^2}{2m}
	\frac{\left.\mathscr{W}_{x'}(\psi_{\epsilon'},\psi_{\epsilon})\right|_0^x}{(\epsilon'-\epsilon)} \;,
\label{eq:normIntW}
\end{equation}
where $\mathscr{W}_{x}(f ,g)$ is a Wronskian with respect to $x$ defined by
\begin{equation}
\mathscr{W}_{x}(f ,g) := f\left(\frac{d}{d x} g\right) - \left(\frac{d}{d x}f\right)g \;.
\end{equation}
With $x=0$ being a pinned point defined by the boundary condition of $\psi_\epsilon|_{x=0}=0$ \textit{independent of energies}, we also have $\psi_{\epsilon'}|_{x=0}=0$, leading to $\mathscr{W}_{x'}(\psi_{\epsilon'},\psi_{\epsilon})|_0 = 0$. Taking the limit of $\Delta\epsilon = \epsilon'-\epsilon \to 0$ on both sides of the equation and using the definition of a partial derivative, we obtain Eq.~(\ref{eq:rgR1D}), from which Eq.~(\ref{eq:rgFR1D}). The derivation of the fundamental equation of rigidity in 3-D, Eq.~(\ref{eq:rgFR3D}), follows a similar script. We note that the same boundary condition, $\psi_\epsilon|_{x=0}=0$, being satisfied at different energies, namely $x=0$ being a special node that does not move with energy, is critical to the derivation, and to the very definition of a pinned point.

The normalization integrals given in forms similar to Eq.~(\ref{eq:normIntW}) or its 3-D equivalent have been well known and used in other contexts especially in QDT formulations \cite{Seaton83,Greene79}. Its implications on the energy derivatives, in the forms of  Eqs.~(\ref{eq:rgR1D}) and (\ref{eq:rgR3D}), were first given in Ref.~\cite{Gao05b} and used for the evaluation of normalization integrals. Further implications of these relations were suspected by us for many years, but never fully understood until much later through the concept of rigidity, as presented here.

\section{Cusp functions and normalization constants for single-term potentials}
\label{sec:fcp1tDeriv}

The cusp functions for single-scale potentials are obtained from the zero-energy regular solution of the radial equation for a single-scale potential $v^{(\alpha)}(r) = \pm D_{\alpha}/r^{\alpha}$ with $\alpha\neq 2$,  
\begin{equation}
\left[-\frac{\hbar^2}{2\mu}\frac{d^2}{dr^2} 
	+ \frac{\hbar^2 \ell(\ell+1)}{2\mu r^2}
	\pm \frac{D_{\alpha}}{r^\alpha} \right]
u^{(\alpha\pm)}_{\epsilon\ell}(r) = \epsilon u^{(\alpha\pm)}_{\epsilon\ell}(r).
\label{eq:rschCusp1s}
\end{equation}
Specifically, the cusp function $f^{cp(\alpha\pm)}_{\ell}(r)$ for potential $v^{(\alpha)}(r)$ is obtained from its zero-energy cusp solution
\begin{equation}
f^{cp(\alpha\pm)}_{\ell}(r) = u^{cp(\alpha\pm)}_{\epsilon=0,\ell}(r) \;,
\end{equation}
satisfying the boundary condition
\[
u^{cp(\alpha\pm)}_{\epsilon=0,\ell}(r) \overset{r\to 0}{\sim} 0 \;,
\]
with an energy-independent normalization constant. Its corresponding strict cusp function is given by
\begin{equation}
F^{cp(\alpha\pm)}_{\ell}(r) = \lim_{r\to 0} f^{cp(\alpha\pm)}_{\ell}(r) \;,
\end{equation}
where $r\to 0$ is, strictly speaking, the strict cusp version of the coalescence limit: $r/\beta_{\alpha}\to 0$.

The ``repulsive'' and ``attractive'' potentials, corresponding the the $\pm D_{\alpha}$, respectively, are treated and labeled separately because their solutions are, in general, qualitatively different. Mathematically, this is reflected as the solution being generally singular at $D_{\alpha}=0$, separating repulsive and attractive solutions. (Rigorously speaking, the true physical meaning of being repulsive and attractive gets reversed for $\alpha<0$. Our simplified naming convention should not lead to any confusion.)

Using a scaled radius, $r_s = r/\beta_{\alpha}$, and making a further change of variable to 
\begin{equation}
y := \frac{2}{|\alpha-2|}r_s^{(2-\alpha)/2} \;,
\end{equation}
to rewrite the zero-energy cusp solution as
\begin{equation}
u^{cp(\alpha\pm)}_{\epsilon=0,\ell}(r) = C_{\ell}r_s^{1/2}w(y) \;,
\end{equation}
one can show that $w(y)$ satisfies
\begin{equation}
\left[ y^2\frac{d^2}{dy^2} + y\frac{d}{dy} \mp y^2-\nu_0^2 \right] w(y) = 0 \;.
\end{equation}
where $\mp$ corresponds to the $\pm$ of Eq.~(\ref{eq:rschCusp1s}), respectively, and
\begin{equation}
\nu_0 := \frac{2}{|\alpha-2|}(\ell+1/2) \;.
\end{equation}
The function $w(y)$ has the Bessel class \cite{Olver10} of linearly independent solutions $J_{\nu_0}(y)$ and $Y_{\nu_0}(y)$ for an ``attractive'' ($-D_{\alpha}$) potential, and linearly independent solutions $I_{\nu_0}(y)$ and $K_{\nu_0}(y)$ for a ``repulsive'' ($+D_{\alpha}$) potential, from which we need only to identify the regular solution and to choose a proper energy-independent normalization constant.

In identifying the regular solution, we note that depending on the value of $\alpha$, whether $\alpha<2$ or $\alpha>2$, the limit of $r\to 0$ means different limits for $y$. For the GC class with $\alpha<2$, the limit of $r\to 0$ corresponds to the limit of $y\to 0$. For the VdW classes with $\alpha>2$, the limit of $r\to 0$ corresponds to the limit of $y\to\infty$. We have thus different cusp functions depending both on the sign of the potential, and on whether $\alpha<2$ or $\alpha>2$.

For the ``attractive'' GC class (aGC), we have
\begin{equation}
f^{cp(\alpha-)}_{\ell}(r) \overset{\alpha<2}{=} 
	b_{\ell}\sqrt{\frac{2}{(2-\alpha)}}r_s^{1/2}J_{\nu_0}(y)\;.
\end{equation}
For the ``repulsive'' GC class (rGC), we have
\begin{equation}
f^{cp(\alpha+)}_{\ell}(r) \overset{\alpha<2}{=} 
	b_{\ell}\sqrt{\frac{2}{(2-\alpha)}}r_s^{1/2}I_{\nu_0}(y)\;.
\end{equation}
Contained in these results are the special cases of $\alpha=1$ for attractive and repulsive Coulomb potentials, the special case of $\alpha=0$ for attractive and repulsive constant potentials, the special case of $\alpha=-1$ for attractive and repulsive linear potentials, and the special case of $\alpha=-2$ for attractive and repulsive quadratic potentials.

For the repulsive van der Waals (rVdW) class, we have 
\begin{equation}
f^{cp(\alpha+)}_{\ell}(r) \overset{\alpha>2}{=} \frac{2}{\pi}\frac{1}{\sqrt{(\alpha-2)}}r_s^{1/2}K_{\nu_0}(y)\;.
\end{equation}
The attractive van der Waals (aVdW) class is nonphysical as a behavior at $r=0$, as discussed in the main text. It does have solution at zero energy. In fact it has a pair of linearly independent solutions that are a part of the QDT formulation for long-range correlation \cite{OMalley61,Gao01,Gao04b}.

The choices of normalization constants for cusp functions are in principle arbitrary. In the main text, we discussed the choice of $b_{\ell}$ for the purpose of grouping together the aGC and rGC solutions and the F class solution. There are other considerations that go into our specific choices. We briefly mention them here since it also helps to put the cusp function within a greater context of a more complete QDT for short-range correlation or more generally a QDT for a reference potential that has a physically-allowed behavior at the origin, instead of a nonphysical behavior such as the aVdW class \cite{Gao08a}.

In a complete QDT for a single-scale potential of $v^{(\alpha)}(r) = \pm D_{\alpha}/r^{\alpha}$ with $\alpha\neq 2$, we would also consider the energy dependence of the solution $u^{(\alpha\pm)}_{\epsilon\ell}$ of Eq.~(\ref{eq:rschCusp1s}), and an irregular solution in addition to the regular solution. In such a theory, we would have a pair of linearly independent solutions, $f^{c(\alpha\pm)}_{\epsilon_s\ell}(r_s)$ and $g^{c(\alpha\pm)}_{\epsilon_s\ell}(r_s)$, at each scaled energy $\epsilon_s = \epsilon/s_E$ where $s_E = \hbar^2/2\mu\beta_{\alpha}^2$, similar to the QDT for the aVdW class \cite{Gao08a}. The cusp function is a special case of $f^{c(\alpha\pm)}_{\epsilon_s\ell}$ at zero energy, namely
\[
f^{cp(\alpha\pm)}_{\ell}(r) = f^{c(\alpha\pm)}_{\epsilon_s=0\ell}(r_s) \;,
\]
and corresponding to each $f^{cp(\alpha\pm)}_{\ell}$, there is an irregular solution
\[
g^{cp(\alpha\pm)}_{\ell}(r) = g^{c(\alpha\pm)}_{\epsilon_s=0\ell}(r_s) \;.
\]
Their normalization constants are chosen such that
\begin{equation}
\mathscr{W}_{r_s}(f^{cp(\alpha\pm)}_{\ell}, g^{cp(\alpha\pm)}_{\ell}) = \frac{2}{\pi} \;,
\label{eq:Wfcpgcp}
\end{equation}
is satisfied, to be consistent with conventions that we have adopted in other QDTs \cite{Gao08a}. Here $\mathscr{W}_{r_s}(f,g)$ is a Wronskian with respect to $r_s$, defined by
\begin{equation}
\mathscr{W}_{r_s}(f ,g) := f\left(\frac{d}{d r_s} g\right) - \left(\frac{d}{d r_s}f\right)g \;.
\end{equation}

While we do not use the irregular solutions, $g^{cp(\alpha\pm)}_{\ell}(r)$, explicitly in this work, we give them here for completeness, and in preparation for future applications. For the ``attractive'' GC class (aGC), we have
\begin{equation}
g^{cp(\alpha-)}_{\ell}(r) \overset{\alpha<2}{=} \frac{1}{b_{\ell}}
	\sqrt{\frac{2}{(2-\alpha)}}r_s^{1/2}Y_{\nu_0}(y)\;.
\end{equation}
For the ``repulsive'' GC class (aGC), we have
\begin{equation}
g^{cp(\alpha+)}_{\ell}(r) \overset{\alpha<2}{=} 
	-\frac{1}{b_{\ell}}\frac{2}{\pi}\sqrt{\frac{2}{(2-\alpha)}}
	r_s^{1/2}K_{\nu_0}(y)\;.
\end{equation}
For the repulsive van der Waals (rVdW) class, we have
\begin{equation}
g^{cp(\alpha+)}_{\ell}(r) \overset{\alpha>2}{=} -\frac{2}{\sqrt{(\alpha-2)}}r_s^{1/2}
	\frac{1}{2}\left[I_{\nu_0}(y) + I_{-\nu_0}(y)\right] \;.
\end{equation}

The normalization constants for the F and alCD classes of solutions are chosen with similar considerations with respect to their corresponding irregular solutions,
\begin{equation}
g^{cp(F)}_{\ell}(r, s_{L}) =  
	-\frac{2^{\ell+\tfrac{1}{2}}\Gamma(\ell+\tfrac{1}{2})}{\pi} (r/s_{L})^{-\ell}\;.
\label{eq:gcpF}
\end{equation}
for the F class, and
\begin{equation}
g^{cp(2)}_{\ell}(r, s_{L}) =  
	-\frac{2^{\ell_t+\tfrac{1}{2}}\Gamma(\ell_t+\tfrac{1}{2})}{\pi} (r/s_{L})^{-\ell_t}\;.
\label{eq:gcpalCD}
\end{equation}
for the alCD class. They are chosen such that Eq.~(\ref{eq:Wfcpgcp}) is satisfied with an understanding of $r_s = r/s_L$ for those classes.

\section{Notes on 2-particle separability}
\label{sec:SepNotes}

The Taylor expansion in 3-D can be written as
\begin{align}
f(\bm{x}+\Delta\bm{x}) &= \exp(\Delta\bm{x}\cdot\nabla)f(\bm{x}) \;,\\
	&= \sum_{j=0}^\infty \frac{1}{j!}(\Delta\bm{x}\cdot\nabla)^j f(\bm{x}) \;.
\end{align}
It gives, for a central potential $v(r)$,
\begin{multline}
v(|\bm{R}+\Delta\bm{R}|) = v(R)+\Delta\bm{R}\cdot\nabla v(R) 
	+ \frac{1}{2}(\Delta\bm{R}\cdot\nabla)^2 v(R) + \dots \;,\\
=  v(R)+\Delta R\ v'(R) P_1(\cos\gamma) \\
	+ \frac{1}{6}(\Delta R)^2 [v''(R)+2 v'(R)/R] \\
	+ \frac{1}{3}(\Delta R)^2 [v''(R) - v'(R)/R]P_2(\cos\gamma) \;,
\end{multline}
where $P_{\ell}(x)$ is the standard Legendre polynomial \cite{Arfken13}, and $\gamma$ is the angle between $\bm{R}$ and $\Delta\bm{R}$.

Applying this result to $v_{ik}(r_{ik})+v_{jk}(r_{jk})$, we obtain
\begin{align}
& v_{ik}(|\bm{r}_k-\bm{r}_i|) + v_{jk}(|\bm{r}_k-\bm{r}_j|) \nonumber\\
&= v_{ik}(R_k) + v_{jk}(R_k) \nonumber\\
& +\frac{m_j}{m_i+m_j}r\ v_{ik}'(R_k) P_1(\cos\gamma) \nonumber\\
& - \frac{m_i}{m_i+m_j}r\ v_{jk}'(R_k) P_1(\cos\gamma) + \dots \;,
\label{eq:Vijdif}
\end{align}
for particles $i$ and $j$ being different and with $R_k := |\bm{r}_k-\bm{c}|$ and $\gamma$ being the angle between $\bm{R}_k := \bm{r}_k-\bm{c}$ and $\bm{r}:=\bm{r}_j-\bm{r}_i$. For particles $i$ and $j$ being identical such that $v_{ik}(r)=v_{jk}(r) =: v(r)$, we obtain
\begin{align}
& v_{ik}(|\bm{r}_k-\bm{r}_i|) + v_{jk}(|\bm{r}_k-\bm{r}_j|) \nonumber\\
&= 2v(R_k) + \frac{1}{12}r^2 [v''(R_k)+2 v'(R_k)/R_k] \nonumber\\
& +\frac{1}{6}r^2 [v''(R_k) - v'(R_k)/R_k]P_2(\cos\gamma) + \dots \;
\label{eq:Vijid}
\end{align}
These results make it clear that the 2-particle separability condition for potential $V^{(ij)}$ [cf. Eq.~(\ref{eq:NbVsep})]
\[
V^{(ij)}\overset{r\to 0}{\sim} V^{(ij)}_{sp} \;,
\]
is satisfied in the generic coalescence limit of $r\to 0$ for any $v_{ik}(r)$ that is continuous in $(0,\infty)$, which we have required for a physical potential in Sec.~\ref{sec:2body}, with only a minor additional requirement of $v_{ik}(r)$  being at least once or twice differentiable ($C^1$ or $C^2$ class).

A deeper understanding of the separability requires a closer look at the correction terms. For an $N$-body system, as discussed in the main text, the order of $R_k$ is measured by an emergent length $r_{\rho}$, which, for a many-body system, can be estimated as $r_{\rho}= (4\pi\rho_{\#}/3)^{-1/3}$ in terms of a local number density $\rho_{\#}(\bm{r})$. Equations~(\ref{eq:Vijdif}) and (\ref{eq:Vijid}) show that $V^{(ij)}-V^{(ij)}_{sp}$ is an expansion in $r v'(r_{\rho})/v(r_{\rho})$, which, to be a small parameter, requires that $v(r)$ is a slowly varying function around $r_{\rho}$ such that the magnitude of its change over $r$, $|r v'(r_{\rho})|$, is small compared to the magnitude of $v(r_{\rho})$ itself. A qualitative understanding of this parameter will help us better understand the more precise meaning of $r\to 0$. Is it $r/\beta_{\alpha}\to 0$, which is strict cusp version of the coalescence limit, or $r/r_{\rho}\to 0$?

A physical interaction potential $v(r)$ can always be characterized by a finite number of length scales, with $\beta_{\alpha}$ denoting the shortest, $\beta_n$ denoting the longest, and a finite number of other scales describing a finite intermediate region, if necessary. For a tightly-bound few-body state or a high-density many-body system, $r_{\rho}\sim\beta_{\alpha}$, and there is no difference between $r/\beta_{\alpha}\to 0$ and $r/r_{\rho}\to 0$. The key is with ``low-density'' states  (which we also use to include loosely-bound few-body states) where $r_{\rho}\gg\beta_{\alpha}$, for which the two limits can be very different.

For interactions of electromagnetic type, such as those between atoms and molecules, $v(r)$ can always be characterized, at sufficiently long range corresponding typically to $r\sim\beta_n$ or greater, by a single-term $v^{(n)} := \pm C_n/r^n$ \cite{Stone13,Gao20b}. At a sufficiently low-density, where $r_{\rho}\sim \beta_n$ or $r_{\rho}> \beta_n$ 
\[
|r v'(r_{\rho})/v(r_{\rho})| = |n|(r/r_{\rho}) \;.
\]
Thus the correction to a separable potential is in this case an expansion in $r/r_{\rho}$, independent of the strength parameter $C_n$ or its corresponding length scale $\beta_n$. The more precise definition of $r\to 0$ that defines the separable limit is in this case $r/r_{\rho}\to 0$, or characterized as an asymptotic region as the region of $r\ll r_{\rho}$.

The distinction between the two versions of the coalescence limit is less significant for smaller $r_{\rho}$ more comparable to $\beta_{\alpha}$, where the potential of electromagnetic type would generally require multiple terms of the form $\pm C_{j}/r^{j}$ or $\pm D_{\alpha}/r^{\alpha}$. But it is still worth noting that even in region of such multi-term potentials, $|r v'(r_{\rho})/v(r_{\rho})| \sim (r/r_{\rho})$ remains true qualitatively because it is true for each term. Thus for interactions of electromagnetic type, the separable limit can very generally be characterized by $r/r_{\rho}\to 0$.

The dual-expansion nature around the coalescence, or equivalently the dual meaning of the coalescence limit of $r\to 0$ as either the strict cusp limit of $r/\beta_{\alpha}\to 0$ or the separable limit of $r/r_{\rho}\to 0$, is an importance characteristic of $(N>2)$-body systems of electromagnetic type. For comparison, if $v(r)$ were of a Yukawa-type nuclear potential resulting from exchange of particles of finite mass, e.g.,  $v(r) = G_Y e^{-r/\beta_Y}/r$ with $\beta_Y$ representing the range of the interaction \cite{Yukawa35}, we would have,
\[
|r\ v'(r_{\rho})/v(r_{\rho})| \sim r/\beta_Y \;,
\]
for $r_{\rho}> \beta_Y$. The corrections to a separable potential would in this case be an expansion in $r/\beta_Y$, and the separable limit would be the same as the strict cusp limit of $r/\beta_Y\to 0$ regardless of density. Thus the emergent length scale, $r_{\rho}$, which exists for all $(N>2)$-body system, does not always mean dual expansion around coalescence, as it does for interactions of electromagnetic type.

\end{document}